\documentclass{article}

\AtBeginDocument{%
	\providecommand\BibTeX{{%
			\normalfont B\kern-0.5em{\scshape i\kern-0.25em b}\kern-0.8em\TeX}}}

\usepackage{mhchem}
\usepackage{cite}
\usepackage{graphicx}
\usepackage{listings}
\usepackage{multirow,rotating}
\usepackage{alltt}
\usepackage{tikz}
\usepackage{algorithm}
\usepackage{float}

\usepackage{soul}

\usepackage{verbatim,afterpage,url,xfrac, cite}
\usepackage[noend]{algpseudocode}

\usepackage{booktabs}
\usepackage{braket}

\long\def\comment#1{}

\def\parah#1{\vspace*{0.0in} \noindent{\bf #1:}}

\newcommand{\cn}{\textsc{CNOT\ }}
\newcommand{\cns}{\textsc{CNOTs\ }}

\newcommand{\linett}{
 
  \coordinate (Y) at (0,0);
  \coordinate (Z) at (0,1);
  \fill[white!10, draw=black] (Y) -- (Z);
 \filldraw (Y) circle (3pt); \filldraw (Z) circle (3pt);
}

\usepackage[sort&compress,square,comma,numbers]{natbib}
\usepackage[colorlinks=true,urlcolor=blue,citecolor=red,linkcolor=red,bookmarks=true]{hyperref}
\usepackage{fullpage,authblk}

\begin{document}

\title{LEAP: Scaling Numerical Optimization Based Synthesis Using an
	Incremental Approach
}

\author[a]{Ethan Smith}
\author[a]{Marc G. Davis}

\author[b]{Jeffrey Larson}

\author[c]{Ed Younis}
\author[c]{Costin Iancu}
\author[c]{Wim Lavrijsen}
\affil[a]{University of California, Berkeley, {\tt ethanhs@berkeley.edu}, {\tt marc.davis@berkeley.edu} }
\affil[b]{Argonne National Laboratory, {\tt jmlarson@anl.gov}}
\affil[c]{Lawrence Berkeley National Laboratory, {\tt edyounis@lbl.gov}, {\tt cciancu@lbl.gov}, {\tt wlavrijsen@lbl.gov} }
\date{}                     
\renewcommand\Affilfont{\itshape\small}

\maketitle
\begin{abstract}
	\it While  showing  great promise, circuit synthesis techniques that combine numerical optimization with search over circuit structures face
	scalability challenges due to a large number of parameters,
	exponential search spaces, and complex objective functions.  The
	LEAP algorithm improves scaling across these dimensions using
	iterative circuit synthesis, incremental
	re-optimization, dimensionality reduction, and improved numerical optimization. LEAP
	draws on the design of the optimal synthesis algorithm QSearch by extending it with an 
	incremental approach to determine constant prefix
	solutions for a circuit. By narrowing the search space, LEAP  improves scalability from four to six
	qubit circuits.  LEAP was evaluated with known quantum circuits
	such as QFT and physical simulation circuits like the VQE, TFIM,
	and QITE. LEAP can compile four qubit
	unitaries up to $59\times$ faster than QSearch and five and six qubit
	unitaries with up to $1.2\times$ fewer CNOTs compared to the  QFAST
	package. LEAP can reduce the CNOT count by up to $36\times$,
	or $7\times$ on average, compared to the CQC Tket
        compiler. Despite its heuristics, LEAP has generated optimal
        circuits for many test cases with a priori known solutions. 
The techniques introduced by LEAP are
	applicable to other numerical-optimization-based synthesis approaches.
\end{abstract}

\section{Introduction}

Quantum synthesis techniques generate circuits from high-level mathematical
descriptions of an algorithm. They can provide a powerful tool for
circuit optimization, hardware design exploration, and algorithm discovery. An
important quality metric of synthesis, and of compilers in general,  is circuit
depth, which relates directly to the program performance on hardware. Short-depth circuits
are especially important for noisy intermediate-scale quantum (NISQ) era
devices, characterized by limited coherence time and
noisy gates. Here synthesis provides a critical capability in enabling experimentation where
 only the shortest depth circuits provide usable outputs.

In general, two concepts are important when thinking about synthesis
algorithms~\citep{raban,synthcsd,ionsynth,qaqc,tucci2005kak,qsearch}:
circuit {\it structure} captures the
application of gates on a ``physical'' qubit link, while {\it function}
captures the gate operations, for example, rotation angle $R_z(\theta)$. Recently
introduced techniques~\citep{qsearch,younis2020qfast}  can generate short-depth circuits in a
topology-aware manner by combining numerical optimization of parameterized
gate representations (e.g., $U_3$) to determine function together with search
over circuit structures.  Regarding circuit depth, their efficacy
surpasses that of traditional optimizing compilers
such as IBM Qiskit~\citep{qiskit} and CQC Tket~\citep{tket}, or of other available synthesis tools such as
UniversalQ\footnote{The UniversalQ algorithms have been recently incorporated
into IBM Qiskit. For brevity, in the rest of this paper we will refer
to it as Qiskit-synth. }~\citep{uq}.

An exemplar of synthesis approaches is QSearch~\citep{qsearch}, which provides optimal-depth
synthesis and has been shown to match known optimal quantum algorithm
implementations for circuits such as QFT~\citep{qft}. QSearch grows a circuit by
adding layers of parameterized gates
and permuting gate placement at each link, building on the
previous best placements to form a circuit {\it structure}. A numerical optimizer
is run on each candidate circuit structure to instantiate the  {\it function}  that
``minimizes'' a score (distance from the target based on the Hilbert--Schmidt norm). This score guides the A*
search algorithm~\citep{astar}  to extend and evaluate the next partial solution.

The QSearch behavior is canonical for numerical-optimization-based
synthesis~\citep{qsearch,ionsynth,qaqc}.
While providing good-quality results, however, these techniques face scalability  challenges:
(1) the number of parameters to optimize grows with circuit depth; (2)  the number of
intermediate solutions to consider is exponential; and (3) the objective
function for optimization is complex, and optimizers may get stuck in local
minima. LEAP (Larger Exploration by Approximate Prefixes) has been designed to improve
the scalability of QSearch, and it introduces several novel techniques
directly extensible to the broader class of search or numerical-optimization-based synthesis.

\parah{Prefix Circuit Synthesis} Designed to improve scaling, LEAP prunes the
search space by limiting  backtracking depth and by coarsening the
granularity of the backtrack steps. Our branch-and-bound algorithm monitors
progress during search and employs ``execution-driven'' heuristics
to decide which partial solutions are good prefix candidates for the final solution. Whenever a
prefix is chosen, the question is whether to reuse the structure (gate
placement) or structure and function (gate instantiation) together. The former approach prunes the search
space, while the latter prunes both the search and parameter spaces.

\parah{Incremental Re-synthesis}  The end result of incremental prefix
synthesis (or
other divide-and-conquer methods, partitioning techniques, etc.) is that circuit pieces
are processed in disjunction, with the potential of missing the global optimum.
Intuitively, LEAP gravitates toward the  solution by combining local
optimization on disjoint sub-circuits. By
chopping and combining pieces of the final circuit, we can create new, unseen
sub-circuits for the optimization process. Overall, this technique is designed
to improve the solution quality for any  divide-and-conquer or other hierarchical approach.

\parah{Dimensionality Reduction} This  technique  could  improve both
scalability and solution quality. QSearch and LEAP require sets of
gates that can fully describe the  Hilbert subspace explored by the
input transformation. This approach ensures convergence,
but in many cases it may overfit the problem. We provide an
algorithm to delete any parameterized gates that do not contribute to the
solution, thereby reducing the dimension of the optimization problems. When
applied directly to the final solution, dimensionality reduction may improve
the solution quality by deleting single-qubit gates. Dimensionality reduction may also
be applied in conjunction with prefix circuit synthesis, improving both
scalability  and solution quality.

 \parah{Multistart Numerical Optimization} This technique  affects
 both scalability and the quality of the solution. Any standalone numerical
 optimizer is likely to have a low success rate when applied to problem
 formulations that involve quantum circuit parameterizations.
 Multistart~\citep{APOSMM} improves on the success rate and quality of solution
 (avoids local minima) by running multiple numerical optimizations  in conjunction.
 Each individual multi-optimization step may become slower, but improved solutions
 may reduce the chance of missing an optimal solution, causing further search expansion. 

LEAP has been implemented as an extension to QSearch, and it has been evaluated
 on traditional ``gates'' such as {\tt mul} and {\tt adder}, as well as 
 full-fledged algorithms such as QFT~\citep{qft}, HLF~\citep{hlf}, VQE~\citep{McClean2015},
 TFIM~\citep{tfimshin,bassman2020domainspecific}, and QITE~\citep{qite}. We compare its behavior with
 state-of-the-art synthesis approaches: QSearch, QFAST~\citep{younis2020qfast}, Tket~\citep{tket}, and
 Qiskit-synth~\citep{uq}.  While QSearch scales up to four qubits, LEAP can 
 compile four-qubit unitaries up to $59\times$ faster than QSearch and scales up
 to six qubits. On well-known quantum circuits such as the Variational Quantum
 Eigensolver (VQE), the Quantum Fourier Transformation (QFT), and physical simulation
 circuits such as the Transverse Field Ising Model (TFIM), LEAP with
 re-synthesis can reduce the \cn count by up to $48\times$, or
 $11\times$ on average.  Our heuristics rarely affect solution quality,  and
 LEAP can frequently match optimal-depth solutions. At five and six qubits,
 LEAP synthesizes circuits with up to $1.19\times$ fewer \cns  on average
 compared with QFAST, albeit
 with  an average $3.55\times$ performance penalty. LEAP can be
 one order of magnitude slower than Qiskit-synth while providing  two or more orders of magnitude shorter
 circuits. Compared with Tket, LEAP reduces the depth on average by $7.70\times$, while taking
  significantly longer in runtime.

All of our techniques affect behavior and performance in a
 nontrivial way:
\begin{itemize}
\item  Compared with QSearch, prefix synthesis reduces by orders of magnitude  the number of  partial
  solutions explored,  leading to significant speedup.
\item Incremental re-synthesis reduces circuit depth by 15\% on average,
  albeit with large increases in  running time.
\item Dimensionality reduction eliminates up to 40\% of $U_3$ gates
  (parameters) and shortens the circuit critical path.
\item Multistart increases the optimizer success rate from 15\% (best value
  observed for any standalone optimizer) to 99\%. For a single optimization run,  however, multistart
  is up to $10\times$ slower than the underlying numerical optimizer.
\end{itemize}

Overall, we believe LEAP provides a very competitive  circuit
optimizer for circuits on NISQ devices up to six qubits. We believe that our techniques can be easily
generalized or transferred directly to other algorithms based on the search of
circuit structures or numerical optimization. For example, re-synthesis, dimensionality reduction, and
multistart are directly applicable to QFAST \textcolor{red}{\textbf{; and re-synthesis is applicable to Qiskit-synth}}.
We can expect that synthesis
techniques using  divide-and-conquer or partitioning methods will be mandatory for
scalability to the number of qubits (in thousands)  provided by future
near-term processors. Our
techniques provide valuable information to these budding approaches.

The rest of this paper is structured as follows. In
Section~\ref{sec:bg} we describe the problem and its challenges. The
proposed solutions are discussed in Sections~\ref{sec:pf}
through~\ref{sec:ms}. The experimental evaluation is presented in
Section~\ref{sec:eval}. In Section~\ref{sec:disc} we discuss the
implications of our approach. Related work is presented in
Section~\ref{sec:related}.
In Section~\ref{sec:conc} we briefly summarize our conclusions.

\section{Background}
\label{sec:bg}

In quantum computing, a qubit is the basic unit of quantum
information. The general
quantum state is represented by a linear combination of two orthonormal basis states (basis vectors).  The most common basis is the equivalent
of the 0 and 1 values used for bits in classical information theory,
respectively {\footnotesize $\ket{0} = \left( \begin{array}{r} 1 \\ 0 \\ \end{array}
\right)$ } and {\footnotesize  $\ket{1} = \left( \begin{array}{r} 0 \\  1
                                  \\  \end{array} \right)$}.

The generic qubit
state is a superposition of the basis states, namely, $\ket{\psi} =
\alpha \ket{0} + \beta \ket{1}$, with complex amplitudes $\alpha$ and
$\beta$, such that $|\alpha|^2+|\beta|^2=1$.

\comment{
$$\left( \matrix{ a & b & c \cr
d & e & f \cr
g & h & i \cr} \right)$$
}

The prevalent model of quantum computation is the circuit model introduced
in~\citep{qcircuit}, where information carried by qubits  (wires) is modified by
quantum gates, which mathematically correspond to unitary operations. A complex
square matrix U is {\bf unitary} if its conjugate transpose $U^*$ is  its
inverse, that is, $UU^* = U^*U = I$.

In the circuit model, a single-qubit gate is represented by a $2 \times 2$
unitary matrix U. The effect of the gate on the qubit state is obtained by
multiplying the U matrix with the vector representing the quantum
state $\ket{\psi'} = U\ket{\psi}$. The most general form of the unitary for a single-qubit
gate is  the ``continuous''
or  ``variational'' gate representation.

\begin{figure}[h]
\centering
{\small
        $U_3(\theta,\phi,\lambda) = \left( \begin{array}{rr}
                                            cos{\frac{\theta}{2}}  & -e^{i\lambda}sin{\frac{\theta}{2}} \\
  e^{i\phi}sin{\frac{\theta}{2}} &
                                   e^{i\lambda+i\phi}cos{\frac{\theta}{2}}
                                            \\ \end{array} \right)$}
\end{figure}

\begin{figure}[htbp!]
 \comment{ [width=3in,height=1in]}
 \centerline{\includegraphics[width=2in,height=.6in,keepaspectratio]{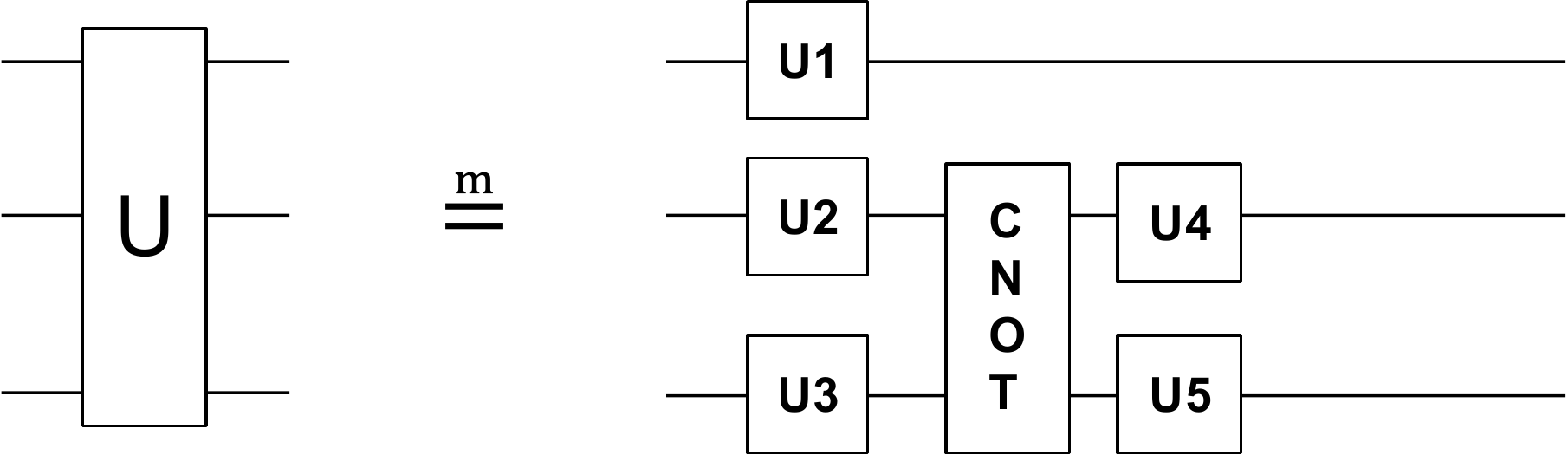}}
 \vspace{.05in}
\centerline{ \includegraphics[width=2.25in,height=.5in,keepaspectratio]{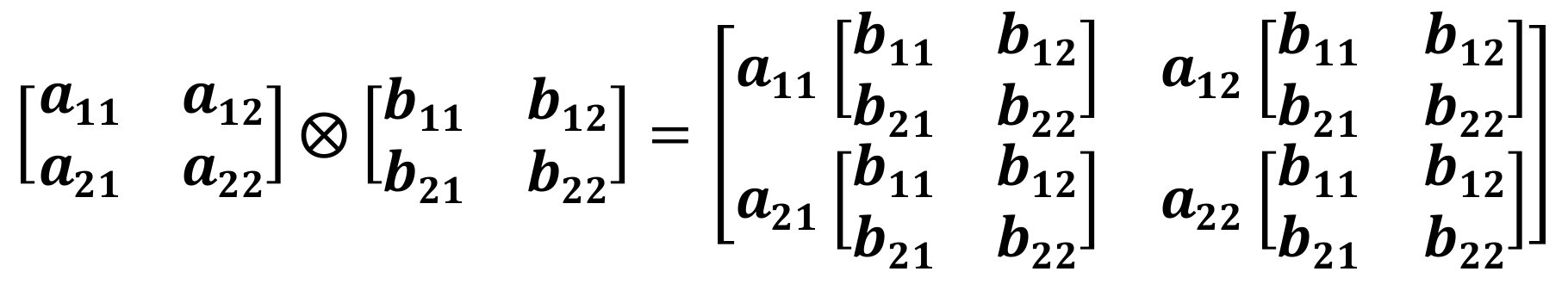}}
  \caption{\label{fig:circ} \it \footnotesize Unitaries (above) and tensors
  products (below). The unitary U represents a $n=3$ qubit transformation,
  where  U is a $2^3 \times 2^3$  matrix. The unitary is
    implemented (equivalent or approximated) by the circuit on the right-hand
    side. The single-qubit unitaries are $2 \times 2$ matrices, while CNOT is a
    $2^2 \times 2^2$ matrix. The computation performed by the circuit is $(I_2
    \otimes U_4 \otimes U_5)(I_2 \otimes CNOT)(U_1 \otimes U_2 \otimes U_3)$, where
    $I_2$ is the identity $2 \times 2$ matrix and $ \otimes $ is the tensor
    product operator. The right-hand side shows the tensor product of $ 2
    \times 2$ matrices.}
 \end{figure}

A quantum transformation (algorithm, circuit)  on $n$ qubits is
represented by a unitary matrix U of size $2^n \times 2^n$.  A circuit is
described by an  evolution in space (application on qubits) and time of gates.
Figure~\ref{fig:circ} shows an example circuit that applies single-qubit and
\cn gates on three qubits.

\parah{Circuit Synthesis} The goal of circuit synthesis is to
decompose unitaries from $SU(n)$ into a product of terms, where each
individual term  (e.g., from $SU(2)$ and $SU(4)$) captures the application of a
quantum gate on individual qubits.  This is depicted in Figure~\ref{fig:circ}.
The quality of a synthesis algorithm is evaluated by the number of
gates in the resulting
circuit and by the  solution distinguishability from the
original unitary.

Circuit length provides one of the main optimality criteria for
synthesis algorithms: shorter circuits are better. \cn count is a
direct indicator of overall circuit length, since the number of 
single-qubit generic gates  introduced in the circuit is proportional to a
constant given by decomposition (e.g., $ZXZXZ$) rules. Since \cn gates
have low fidelity on NISQ devices, state-of-the-art
approaches~\citep{raban,synthcsd} directly attempt to minimize their
count. Longer-term, single-qubit gate count (and circuit critical
path) is likely to augment the
quality metric for synthesis.

Synthesis algorithms use distance metrics  to assess the solution quality. 
Their goal is to minimize $\|U - U_S\|$, where $U$ is the unitary
that describes the transformation and $U_S$ is the computed
solution. They choose an error threshold $\epsilon$ and use it for
convergence, $\| U-U_S \| \le \epsilon$. Early synthesis algorithms used the
diamond norm, while more recent efforts~\citep{qaqc,HSnormsynth} use a metric
based on the Hilbert--Schmidt inner product between $U$ and $U_S$.
\begin{equation} \label{eq:hsn}
\langle U, U_S \rangle_{HS} = Tr(U^{\dag} U_S)
\end{equation}
This is motivated by its lower computational overhead.

 \begin{figure*}
 \centerline{\includegraphics[keepaspectratio=true,width=5in]{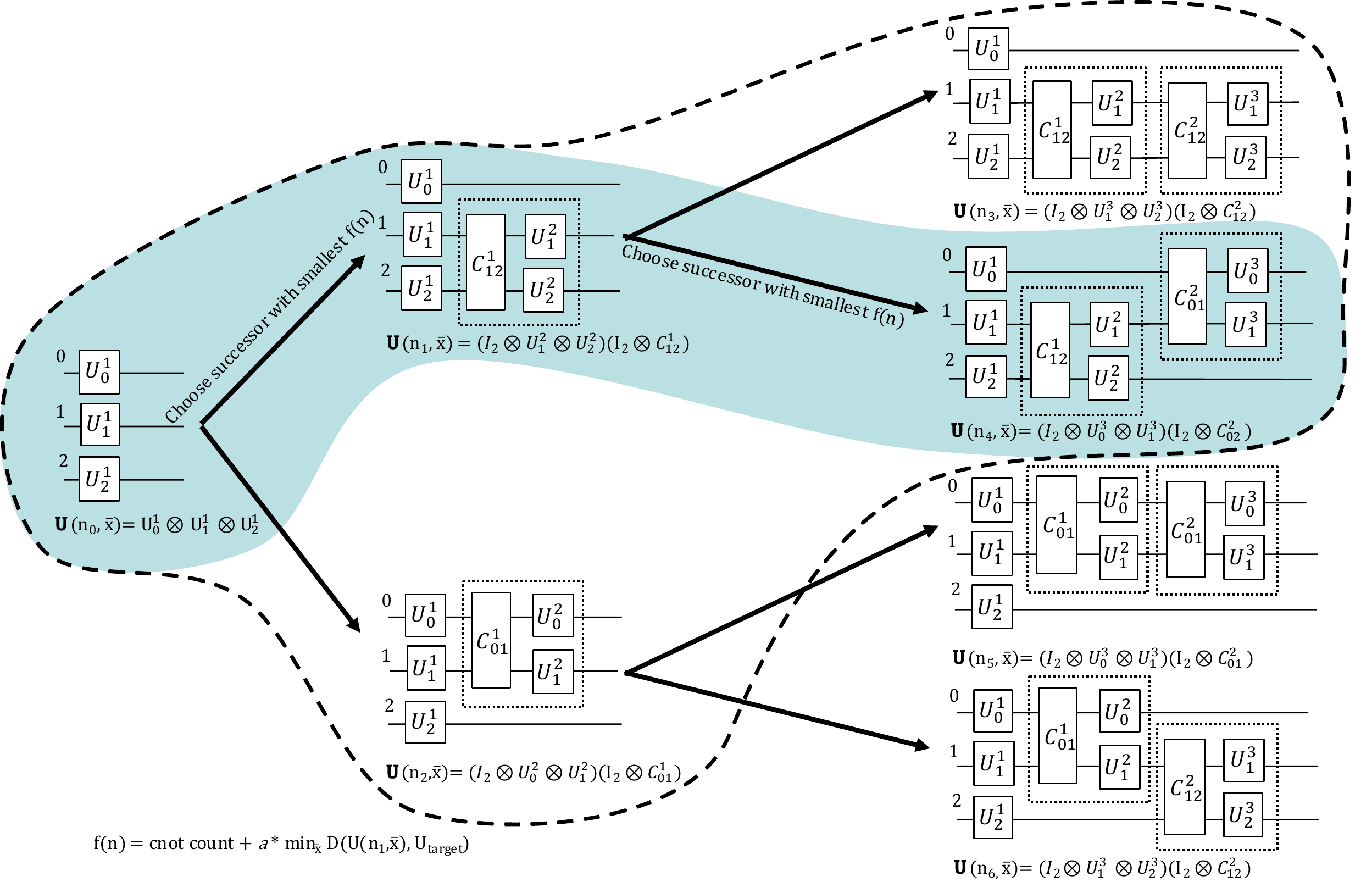}}
\caption{\label{fig:usearch} \footnotesize \it  Example evolution
    of the search algorithm for a three-qubit circuit. It starts by placing a layer of 
    single-qubit gates, then generating the next two possible solutions. Each
    is evaluated, and in this case the upper circuit is closer to the
    target unitary, leading to a smaller heuristic value. This circuit is then expanded with its possible
    two successors. These are again instantiated by the optimizer. The
    second circuit from the top has an acceptable distance and is
    reported as the solution. The path in blue  shows the evolution of
    the solution. The ansatz circuits enclosed by the dotted line have been
    evaluated during the search.}
\end{figure*}

 \subsection{Optimal-Depth Topology-Aware Synthesis}

QSearch~\citep{qsearch} introduces an optimal-depth topology-aware synthesis
algorithm that has been demonstrated to be extensible across native
gate sets (e.g., \{$R_X, R_Z, CNOT$\},  \{$R_X, R_Z,$ \textit{SWAP}\}) and to
multilevel systems such as qutrits.

The approach employed in QSearch is canonical for the operation of
other synthesis approaches that employ numerical
optimization. Conceptually, the problem can be thought of as a
search over a tree of possible circuit structures containing
parameterized gates. A search algorithm
provides a principled way to walk the tree and evaluate candidate
solutions. For each candidate, a numerical optimizer instantiates the
function (parameters) of each gate in order to minimize some distance
objective function. 
 
QSearch works by extending the circuit structure a layer at a time. At each step,
the algorithm places a two-qubit expansion operator in all legal placements. The
operator contains one CNOT gate and two  $U_3(\theta,\phi,\lambda)$ gates. 
QSearch then evaluates these candidates using numerical optimization to
instantiate {\it all} the single-qubit gates in the structure. An
A*~\citep{astar} heuristic determines which of the candidates is
selected for another layer expansion, as well as the destination of
the backtracking steps. Figure~\ref{fig:usearch} illustrates this process for
a three-qubit circuit.

Although theoretically able to solve for any ``program'' (unitary) size,
the scalability of QSearch is limited in practice to four-qubit programs
because of several factors. 
The A* strategy determines the number of solutions evaluated: at best this
is linear in depth; at worst it is exponential. \comment{Our examination of
QSearch performance indicates that its scalability is limited to four
qubits first due to the presence of too many deep backtracking
chains.}  Any technique to reduce the number of candidates, especially
when deep, is likely to improve performance. Our prefix synthesis
solution is discussed 
in Section~\ref{sec:pf}.

Since each
expansion operator has two $U_3$ gates, accounting for six\footnote{In
  practice, QSearch uses 5 parameters because of commutativity rules
  between single-qubit and
  CNOT gates.} parameters, circuit
parameterization grows linearly with depth. Numerical
optimizers scale at best with a high-degree polynomial in the number
of parameters,
making optimization of long circuits challenging. Any
technique to reduce the number of parameters is likely to improve
performance.
Dimensionality reduction is discussed further in Section~\ref{sec:dr}.

The scalability and the quality of the numerical optimizer
matter. Faster optimizers are desirable, but their quality affects performance nontrivially. Our experimentation
with CMA-ES~\citep{cmaes}, L-BFGS~\citep{lbfgs}, and Google
Ceres~\citep{ceres} shows that the QSearch success rate of obtaining a
solution from a valid structure can vary from 20\% to 1\%  for longer circuits. Besides this
measurable outcome, the propensity of optimizers to get stuck in local
minima and plateaus can have an insidious effect on scalability by altering the
search path. A more nuanced approach to optimization and judicious
allocation of optimization time budget may improve scalability. Our
multistart approach
is discussed further in Section~\ref{sec:ms}.

\section{Prefix Circuit Synthesis}
\label{sec:pf}

The synthesis solution space can be thought of as a tree that enumerates
circuit structures of increasing depth: Level 1 contains  depth-one structures,
Level 2 contains  depth-two structures, and so on. For scalability, we want to
reach a solution while evaluating the least number of candidates possible {\it
and} the shallowest circuits possible.  The number of evaluations is given by
the search algorithm: in the case of QSearch the path  is driven by A*,
and scalability is limited by long backtracking chains.

\begin{figure}[h!]
 \comment{ [width=4in,height=1in]}
 \centerline{\includegraphics[width=4in,height=1.3in,keepaspectratio]{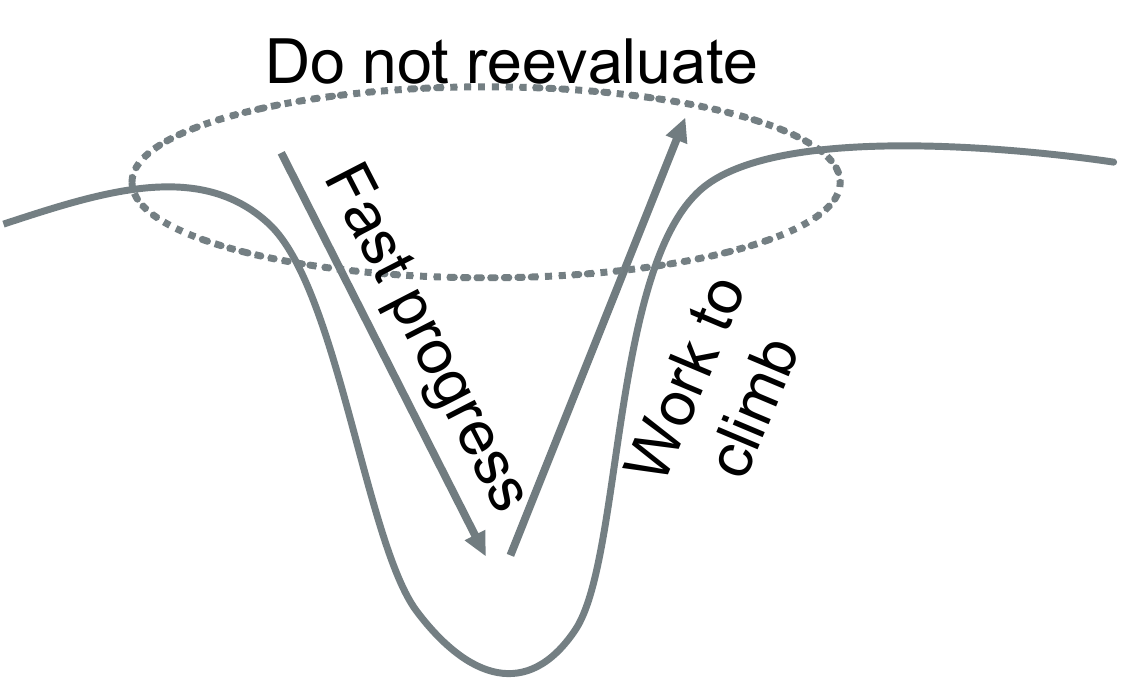}}

  \caption{\label{fig:plateau} \it \footnotesize Synthesis needs to navigate around local minima and plateaus.}
\end{figure}

\begin{figure*}[h!]

  \centerline{\includegraphics[width=5.5in,keepaspectratio]{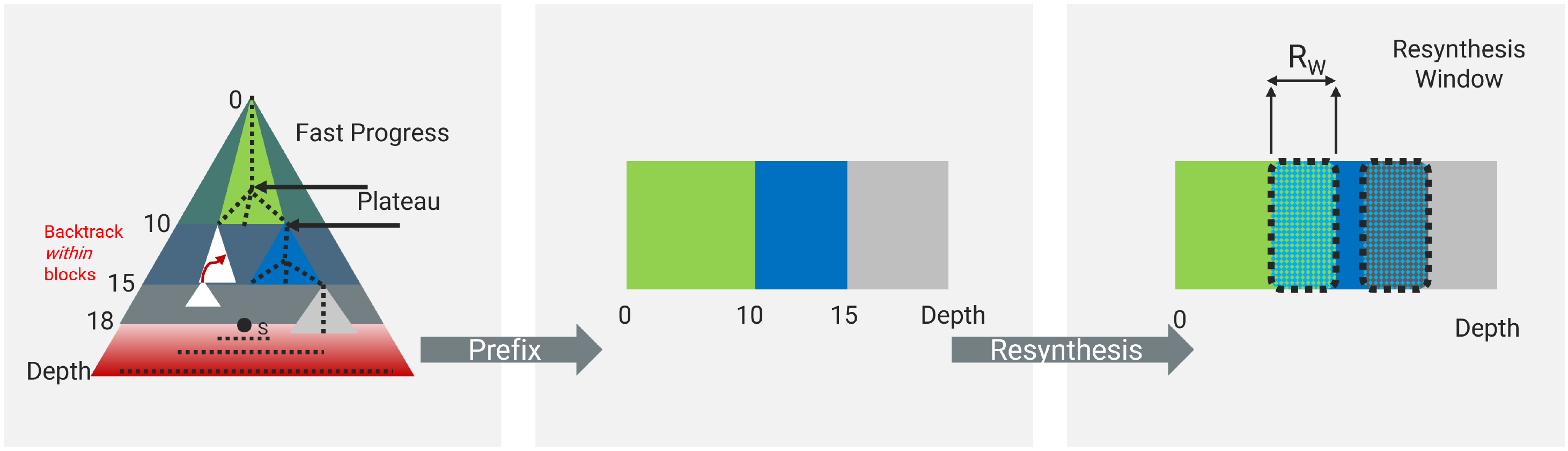}}
 
 \caption{\label{fig:flow} \it \footnotesize Prefix-based synthesis
   induces a partitioning of the circuit. Each partition/prefix
   captures the effect of its associated sub-tree on the search for
   a solution. Each partition has been
   subject to optimization: global with respect to the partition
   itself, but local with respect to the final solution. The resulting
 circuit in the middle reaches a solution from composing local
 optima.  Re-synthesis combines disjoint partitions in order to form
 regions that are passed through optimization. Since the new regions have
 not been subject to optimization, there exists the potential for
 improvement. }
\end{figure*}

Our idea introduces a simple heuristic to reduce the frequency of backtracking. The approach is ``data driven'' and inspired by techniques employed in numerical optimization, as shown in Figure~\ref{fig:plateau}. Imagine mapping the search tree onto an optimization surface, which will contain plateaus and local minima. Exiting a plateau is characterized by faster progress toward a solution and minima. If the minima are local
(partial solution is not acceptable), the algorithm has to walk out of the ``valley.'' Once out, the algorithm may still be on  a plateau, but it can mark the region just explored as not ``interesting'' for any backtracking.    The effect of implementing these principles in the search is illustrated in Figure~\ref{fig:flow}. The result is a partitioning of  the solution space into coarse-grained regions grouped by circuit depth range.  During search, backtracking between solutions within a region is performed by using the A* rules. We never backtrack outside of a region to any candidate solution that resides in the previous ``depth band.''

Overall, the effect of our strategy can be thought of as determining a prefix structure on the resulting circuit, as shown in Figure~\ref{fig:flow}. The algorithm starts with a pure A* search on circuits
up to depth $d_1$. The first depth $d_1$ viable partial solution is recorded, and the search proceeds to depth $d_2$ in sub-tree A. A* search proceeds in sub-tree A until finding the first viable candidate at depth $d_2$, then proceeds in sub-tree B. At this point we have three regions: the start  sub-tree for depth 0 to $d_1$, A for depth $d_1 + 1$ to $d_2$,  and B for depth $d_2 + 1$ to $d_3$. In this example the search in sub-tree B fails at depth $d_2 + 1$.
We, therefore, backtrack to $d_2$, and the search proceeds on the path depicted on the right-hand side of the tree and eventually finds a solution.

One can easily see how by prohibiting backtracking into large solution sub-trees
we can reduce the number of evaluated (numerically optimized) candidates and
improve scalability. As this changes the A* optimality property of the
algorithm, the challenge is determining these sub-trees in a manner that still
leads to a short-depth solution.

\parah{Prefix Formation} A partial solution describes a circuit structure and its function (gates). 
We have considered both static and dynamic methods for prefix formation. In our nomenclature, a static approach will choose a prefix circuit whose structure and function are fixed: this is a fully instantiated circuit. A dynamic approach will choose a fixed structure whose function is still parameterized.  In the first case, the prefix circuit is completely instantiated with native gates to perform a single computation, while in the latter it can ``walk'' a much larger Hilbert subspace as induced by the parameterization.
Intuitively, determining a single instantiated prefix circuit is good for scalability. This reduces the number of parameters evaluated in any numerical optimization operation after prefix formation. We  have experimented with several strategies for forming instantiated prefix circuits in our synthesis algorithms, but they  did not converge or they produced very long circuits.

\parah{Prefix Formulation} In LEAP we use a dynamic data-driven approach
informed by the evolution of the underlying A* QSearch algorithm, described in
Figure~\ref{fig:flow}.  Our analysis of the trajectories for multiple examples
shows that many  paths are characterized by a rapid improvement in solution
quality (reduction in Hilbert--Schmidt distance between target unitary and
approximate prefix), followed by plateauing induced  either by optimizer
limitations (local minima) or as  an artifact of the particular structures
considered (dead-end).

LEAP forms subtrees by first identifying and monitoring plateaus. Since during
a plateau the rate of solution quality change is ``low,''  a ``prefix'' is
formed whenever a solution is evaluated with a jump in the rate of change. The
plateau identification heuristic is augmented with a work-based heuristic: we
wait to form a prefix until we sample enough partial solutions on a path. This
serves several purposes: it gives us more samples in a sub-tree to gain some
confidence we have not skipped ``the only few viable partial solutions,'' and
it increases the backtracking granularity by identifying larger subtrees. Even
more subtly, the work heuristic decreases the sensitivity of the approach to
the thresholds used to assess the rate of change in the plateau identification
method. By delaying to form a prefix based on work, we avoid jumping directly
into another plateau that will result in superfluously evaluating many
solutions that are close in depth to each other.

\parah{Solution Optimality} By discarding pure A* search, LEAP gives up on
always finding the optimal solution. However, the following observations based
on the properties of the solution search space indicate that optimality loss
could be small and that the approach can be generalized to other search and
numerical optimization-based methods.

First, the solution tree of circuit structures exhibits high symmetry. Partial
solutions can be made equivalent by qubit relabeling; all  solutions reached
from any equivalent structure will have a similar depth. For example, for a
circuit with N qubits, a depth 1 circuit with a CNOT on qubits 0 and 1 can be
thought of as ``equivalent'' to the circuit with a CNOT on qubits $N-2$ and $N-1$.
Symmetry indicates that coarse-grained pruning  may be feasible, since a
sub-tree may contain many ``equivalent'' partial solutions.

Second, assuming that the optimal solution has depth $d$, there are many
easy-to-find solutions at $depth > d$. In Figure 3, assume that the solution
node $S$ at depth $d$ is missed by our strategy. However, there are $links$
solutions at $d+1$, $links^2$ solutions at $d+2$, and so forth, trivially
obtained by adding identity gates to $S$. In other words, the solution density
increases (probably quadratically) with circuit depth increase. If the search
has a ``decent'' partial solution at depth $d$, numerical optimization is
likely to find the final solution at very close depth. Overall, the high-level
heuristic goal is to get to optimal depth with a ``good enough'' partial
solution.  Our ``good enough'' criteria combine the Hilbert--Schmidt norm with
a measure of work.

The pseudocode for the prefix formation algorithm in LEAP is presented in Figure~\ref{fig:alg}.

\newcommand{\ui}{U_{implemented} }
\newcommand{\uis}{$U_{implemented}$ }
\newcommand{\ut}{U_{target} }
\newcommand{\uts}{$U_{target}$}

{\footnotesize
\begin{figure*}[ht]
	\centering
	\setlength{\parindent}{0cm}
	\begin{minipage}[t]{0.45\linewidth}
		\begin{algorithm}[H]
			\scriptsize
			\caption{ Helper Functions}
			\begin{algorithmic}[1]
				\Function{s}{$n$}
				\State \textbf{return} $\{n + \cn + U_3 \otimes U_3 \text{ for all possible \cn positions}\}$
				\EndFunction
				\\
				\Function{p}{$n$, $U$}
				\State \textbf{return} ${min}_{\overline{x}} D(U(n, \overline{x}), U)$
				\EndFunction
				\\
				\Function{h}{$d$}
				\State \textbf{return} $d * a$ \Comment{$a$ is a constant determined via experiment. See section 3.3.1}
				\EndFunction
				\\
				\Function{predict\_score}{$a$, $b$, $d_i$}
				\State \textbf{return} \{Predicted \cns for depth $d_i$based on points in a, b\}
				\EndFunction
			\end{algorithmic}
		\end{algorithm}
	\end{minipage}
	\hspace{0.01\linewidth}%
	\begin{minipage}[t]{0.45\linewidth}
		\begin{algorithm}[H]
			\scriptsize
			\caption{LEAP Prefix Formation}
			\begin{algorithmic}[1]
				\Function{leap\_synthesize}{\uts, $\epsilon$, $\delta$}
				\State $s_i \gets$ the best score of prefixes
				\State $n_i \gets$ the prefix structure
				\While{$s_i > \epsilon$}
				\State $n_i, s_i \gets$ \Call{inner\_synthesize}{\uts, $\epsilon$, $\delta$}
				\EndWhile
				\State \textbf{return }$n_i, s_i$
				\EndFunction
				\\
				\Function{inner\_synthesize}{\uts, $\epsilon$, $\delta$}
				\State $n \gets \text{representation of $U_3$ on each qubit}$
				\State $a \gets \text{best depth values of intermediate results}$
				\State $b \gets \text{best depth values of intermediate results}$
				\State \textbf{push } $n$ \textbf{ onto } $queue$ \textbf{ with priority } \Call{h}{$d_{best}$}$ + 0$
				\While{$queue$ is not empty}
				\State $n \gets \textbf{pop from } queue$
				\ForAll{$n_i \in $\Call{s}{$n$}}
				\State $s_i \gets$ \Call{p}{$n_i$, \uts}
				\State $d_i \gets$ \cn count of $n_i$
				\State $s_p \gets$ \Call{predict\_score}{$a$, $b$, $d_i$}
				\If{$s_i < \epsilon$}
				\State \textbf{return }$n_i, s_i$
				\EndIf
				\If{$s_i < s_p$}
				\State \textbf{return }$n_i, s_i$
				\EndIf
				\If {$d_i < \delta$}
				\State \textbf{push } $n_i$ \textbf{ onto } $queue$ \textbf{ with priority } \Call{h}{$d_i$}$ + $\cn count of $n_i$
				\EndIf
				\EndFor
				\EndWhile
				\EndFunction
			\end{algorithmic}
		\end{algorithm}		
              \end{minipage}
              \caption{\label{fig:alg} \it \footnotesize Prefix formation algorithm in LEAP, based on the algorithm in \citep{qsearch}.}
\end{figure*}
}

\section{Incremental Re-synthesis}
\label{sec:ir}

The end result of incremental synthesis (or
other divide-and-conquer methods, partitioning techniques, etc.) is that
circuit pieces are optimized in disjunction, with the potential of
missing the optimal solution. For LEAP, this is illustrated in
Figure~\ref{fig:flow}. Prefix synthesis generates a natural
partitioning of the circuit. Each partition is optimized based on
knowledge local to its sub-tree. The final solution is composed of
local optima. 
The basic observation here is that by chopping and combining pieces of the circuit generated by prefix synthesis, we
can create new, unseen circuits for the optimization process.

For incremental re-synthesis, we use the output circuit from prefix
synthesis and its partitioning (the list of depths
where prefixes were fixed). The reoptimizer
removes circuit segments to create ``holes'' of a size provided by the
user (referred to as re-synthesis window) centered on the divisions
between partitions. This circuit is lifted to a unitary, and
the reoptimizer synthesizes it and replaces it into the original
solution. The process continues iteratively until a stopping criterion
is reached.
This amounts to moving a sliding optimization window across the
circuit.  

The quality of the solution is determined by the choice of the size of
the re-synthesis window, the number of applications (circuit
coverage) and stopping criteria, and the numerical optimizer.

In LEAP we make several pragmatic choices. The size of the
optimization window is selected to be long enough for reduction
potential but overall short enough that it can be optimized fast
enough. The algorithm reoptimizes exactly once at each
boundary in the  original partitioning. The re-synthesis pass allows
us to manage the budget given to numerical optimizers. Since each circuit
piece is likely to be transformed multiple times, some of the
operations can use fast but lower-quality/budget optimization. We do
use the fastest optimizer available during prefix synthesis, switching
during re-synthesis to the higher-quality but slower multistart solver based
on~\citep{APOSMM}, described in Section~\ref{sec:ms}.

\section{Dimensionality Reduction}
\label{sec:dr}

The circuit solution provides a parameterized structure instantiated
for the solution. This parameterization introduced by the single-qubit
$U_3$ gates may overfit the problem.

For LEAP, which targets only the CNOT count, this may be a valid
concern, and we therefore designed a dimensionality reduction pass. We use a
simple algorithm that attempts to delete one $U_3$ gate at a time and reinstantiates the
circuit at each step. This linear complexity algorithm can
discover and remove only simple correlations between parameters. More
complex cases can be discovered borrowing from techniques for
dimensionality reduction for machine learning~\citep{drml} or numerical
optimization~\citep{dropt}.

When applied to the final synthesis solution, dimensionality reduction 
may reduce  the circuit critical path even further by
deleting $U_3$ gates. It can also also be combined with
the prefix synthesis. Once a prefix is formed, we can reduce its
dimensionality. As numerical optimizers scale exponentially with
parameters, this will improve the execution time per invocation. On
the other hand, it may affect the quality of the solution as we remove
expressive power from prefixes. In the current LEAP version, only the final solution is simplified.

\section{Multistart Optimization}
\label{sec:ms}

Solving the optimization problem for the objective function in LEAP or
QSearch can be difficult.
Quantum circuits, even optimal ones, are not unique: a global phase is
physically irrelevant and thus does not affect the output.
Furthermore, circuits that differ only in a local basis transformation and its
inverse surrounding a circuit subsection (e.g., a single 2-qubit gate) are
mathematically equivalent.\footnote{There are physical differences; in
particular such circuits tend to sample different noise profiles.
This property forms the basis of randomized compilation.}
Provided native gate sets may contain equivalences; and single-qubit
gates, being rotations, are periodic.
As a practical matter, we find that we cannot declare these
equivalences to existing optimizers. Furthermore, where they can be used to create
constraints or inaccessible regions (e.g., by remapping the periodicity into
a single region), we find that they hinder the search, because
boundaries can create artificial local minima.

\begin{figure}[htbp!]
 \centerline{\includegraphics[width=2in,keepaspectratio]{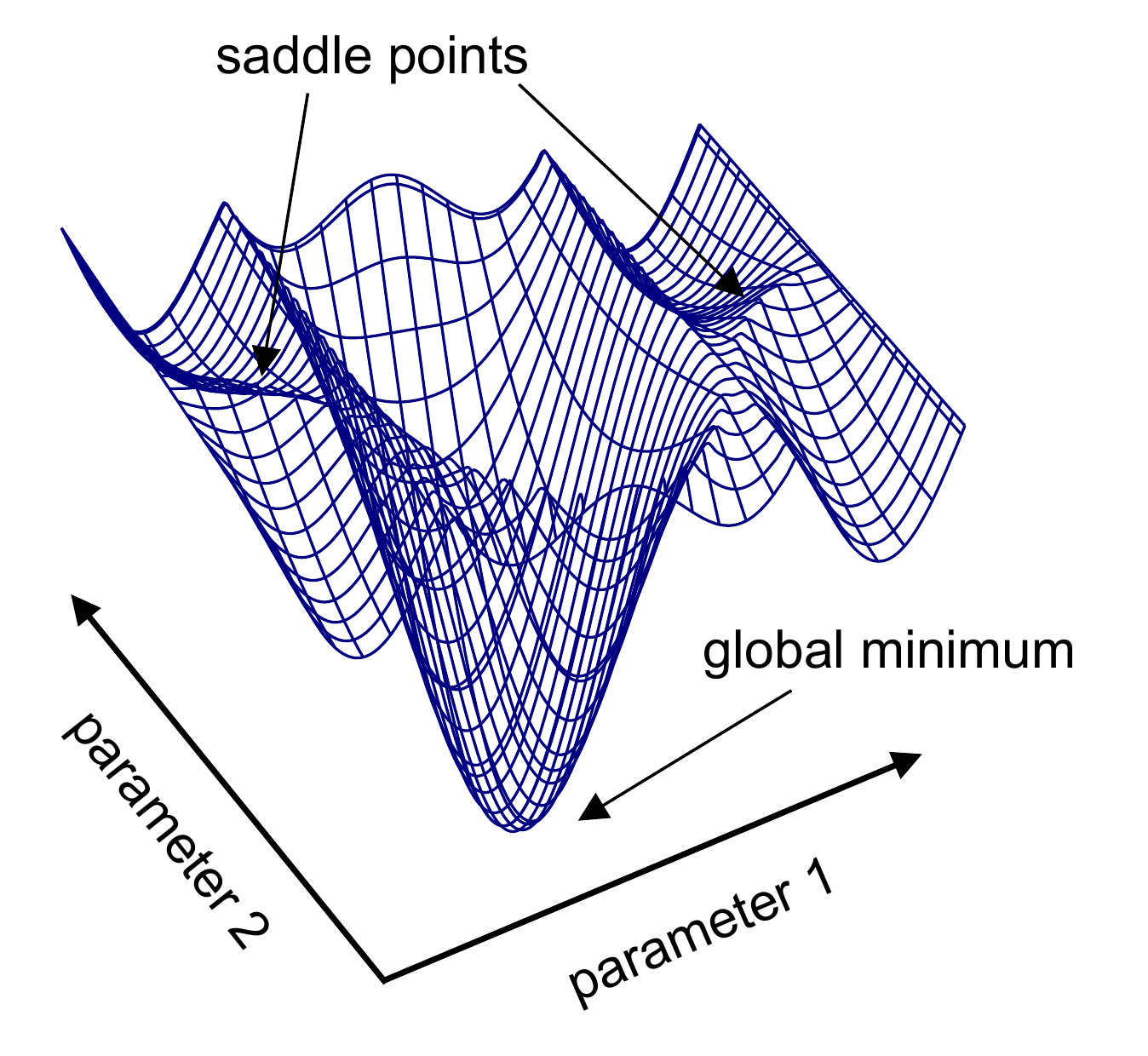}}
  \caption{\label{fig:summa} \it \footnotesize  Optimization surface near the
    global minimum for a 4-qubit circuit of depth 6 for the first step in the
    QITE algorithm, varying 6 (3 pairs) out 42 parameters equally, showing
    the effect on the optimization surface for 2 parameters from distinct
    pairs. (The global minimum is so pronounced only because the remaining 36
    parameters are kept fixed at optimum, reducing the total search space;
    most of the 42-dim surface is flat. \label{fig:qite_surface}}
\end{figure}

The unavoidable presence of equivalent circuits means that we are essentially
overfitting the problem, where changes in parameters can cancel each other out,
leading to saddle points, which turn into local minima in the optimization
surface because of the periodicity; see Figure~\ref{fig:qite_surface}.
The former cause, at best, an increase in the number of iterations as
progress slows down because of smaller gradients; the latter risks getting  the optimizer
stuck.

Another problem comes from the specification of the objective: distance
metrics care only about the output, and different circuits can thus result in
equal distances from the desired unitary.
If no derivatives are available, this results in costly evaluations just to
determine no progress can be made, a problem that gets worse at scale.
But even with a derivative, it closes directions for exploration and shrinks
viable step sizes, thus increasing the likelihood of getting stuck in a local
minimum.

In sum, local optimization methods are highly dependent on the starting
parameters, yet global optimization methods can require far too many
evaluations to be feasible for real-world objectives.
An attractive middle ground is an approach that starts many local optimization runs
from different points in the domain. Multistart
optimization methods are especially appealing when there is some structure in the
objective, such as the least-squares form of the objective. \comment{There exists
local optimization approaches such as \citep{X} or \citep{Y} that
exploit such structure to improve convergence to a stationary point. Using these
approaches within a multistart framework can often greatly improve the final
solution identified.}

Some multistart approaches complete a given local optimization run before
starting another, whereas others may interleave points from different runs.
The asynchronously parallel optimization solver for finding multiple minima
(APOSMM)~\citep{APOSMM} begins with a uniform sampling of the domain and then starts local
optimization runs from any point subject to constraints: (1)
point not yet explored; (2) not a local
optimum; and (3) no point available within a distance $r_k$ with a
smaller function value. If no such point is available, more sampling is
performed. The radius $r_k$ decreases as more points are sampled, thereby
allowing past points to start runs. Under certain conditions on the objective
function and the local optimization method, the logic of APOSMM can be shown to
asymptotically identify all local optima while starting only finitely many local
optimization runs.

\section{Experimental Setup}
\label{sec:eval}

LEAP, available at  {\tt \footnotesize
  https://github.com/BQSKit/qsearch}, extends QSearch. 
We evaluated it with {\tt Python} 3.8.5, using {\tt numpy} 1.19.5 and  {\tt
  Rust} 1.48.0 code.

For our APOSMM implementation, we integrated with the version in the libEnsemble
Python package~\citep{libEnsemble, Hudson2021}. We tried two different local
optimization methods within APOSMM:
the L-BFGS implementation within SciPy~\citep{lbfgs} and the Google Ceres~\citep{ceres} least-squares 
optimization routine.

For experimental evaluation, we use a 3.35GHz Epyc 7702p based server, with 64 cores and 128 threads. Our workload
consists of known circuits (e.g., {\tt mul}, {\tt add}, Quantum
Fourier Transform), as well as newly introduced
algorithms. 
VQE~\citep{McClean2015} starts with a parameterized
circuit and implements a hybrid algorithm where parameters are
reinstantiated based on the results of the previous run. The
TFIM~\citep{tfimshin} and Quantum
Imaginary Time Evolution (QITE)~\citep{qite} algorithms model the time
evolution of a system. They are particularly challenging for NISQ
devices as circuit length grows linearly with the simulated time
step. In TFIM, each timestep (extension) can be computed and compiled ahead of time
from first principles, while in QITE it is dependent on the previous
time step.

We evaluate LEAP against QSearch and other available state-of-the-art
synthesis software and compilers. QFAST~\citep{qfast} scales better
than QSearch by conflating search for structure with numerical
optimization, albeit producing longer circuits. Qiskit-synth~\citep{uq}
uses linear algebra decomposition rules for fast synthesis, but
circuits tend to be long. IBM Qiskit~\citep{qiskit} 
provides ``traditional'' quantum compilation infrastructures using
peephole optimization and mapping algorithms. CQC Tket~\citep{tket}  proves another
good quality compilation infrastructure across multiple gate sets. To showcase the impact of QPU
topology, we compile for processors where qubits are fully connected
(all-to-all), as well as processors with qubits connected in a
nearest-neighbor (linear) fashion.

\begin{table*}[!htp]\centering
    \caption{\it \footnotesize Results for 3-4 qubit synthesis benchmarks. * 3 Qubit results were chosen as the best run of two samples.
    \label{tab:34q}}
    \tiny
    \resizebox{\columnwidth}{!}{%
        \setlength\tabcolsep{3pt}
        \begin{tabular}{|c|c|c|ccccccc|cccccccc|}\toprule
        \textbf{} &\textbf{} &\textbf{} &\multicolumn{7}{c}{\textbf{3 Qubits*}} &\multicolumn{8}{c}{\textbf{4 Qubits}} \\\cmidrule{4-18}
        \textbf{} &\textbf{} &\textbf{} &\textbf{fredkin} &\textbf{toffoli} &\textbf{grover} &\textbf{hhl} &\textbf{or} &\textbf{peres} &\textbf{qft3} &\textbf{adder} &\textbf{vqe} &\textbf{TFIM-1} &\textbf{TFIM-10} &\textbf{TFIM-22} &\textbf{TFIM-60} &\textbf{TFIM-80} &\textbf{TFIM-95} \\\midrule
        \multirow{8}{*}{\textbf{CNOTs}} &\multirow{4}{*}{\textbf{All-to-All}} &\textbf{Qiskit Mapped} &8 &6 &7 &5 &6 &5 &6 &10 &76 &6 &60 &132 &360 &480 &570 \\
        & &\textbf{QFAST} &8 &8 &7 &3 &8 &7 &7 &15 &43 &8 &14 &16 &18 &14 &21 \\
        & &\textbf{LEAP} &7 &6 &6 &3 &6 &5 &6 &12 &22 &6 &12 &13 &12 &15 &12 \\
        & &\textbf{TKet Mapped} &8 &6 &7 &3 &6 &5 &6 &10 &71 &6 &60 &132 &360 &480 &570 \\
        & &\textbf{Qiskit Synthesized} &15 &9 &29 &13 &11 &11 &27 &66 &566 &124 &218 &218 &218 &218 &218 \\\cmidrule{2-18}
        &\multirow{4}{*}{\textbf{Linear}} &\textbf{Qiskit Mapped} &12 &13 &14 &11 &11 &9 &8 &20 &85 &6 &60 &132 &360 &480 &570 \\
        & &\textbf{QFAST} &8 &8 &7 &4 &8 &7 &8 &36 &40 &6 &10 &10 &12 &12 &23 \\
        & &\textbf{LEAP} &8 &8 &7 &3 &8 &7 &7 &14 &24 &7 &12 &13 &12 &13 &12 \\
        & &\textbf{TKet Mapped} &14 &9 &13 &3 &12 &11 &9 &16 &71 &6 &60 &132 &360 &480 &570 \\
        & &\textbf{Qiskit Synthesized} &30 &17 &74 &30 &19 &28 &70 &247 &2630 &477 &523 &523 &523 &523 &523 \\\cmidrule{1-18}
        \multirow{8}{*}{$U_3$\textbf{s}} &\multirow{4}{*}{\textbf{All-to-All}} &\textbf{Qiskit Mapped} &10 &8 &17 &10 &9 &9 &11 &11 &86 &7 &70 &154 &420 &560 &665 \\
        & &\textbf{QFAST} &19 &19 &17 &9 &19 &17 &17 &34 &91 &20 &32 &36 &40 &32 &46 \\
        & &\textbf{LEAP} &17 &15 &15 &9 &15 &13 &15 &26 &49 &16 &28 &28 &28 &31 &28 \\
        & &\textbf{TKet Mapped} &10 &8 &16 &5 &8 &9 &11 &10 &76 &7 &61 &133 &361 &481 &571 \\
        & &\textbf{Qiskit Synthesized} &19 &11 &42 &17 &17 &12 &39 &88 &671 &160 &261 &261 &261 &261 &261 \\\cmidrule{2-18}
        &\multirow{4}{*}{\textbf{Linear}} &\textbf{Qiskit Mapped} &23 &22 &30 &22 &22 &20 &18 &37 &106 &7 &70 &154 &420 &560 &665 \\
        & &\textbf{QFAST} &19 &19 &17 &11 &19 &17 &19 &76 &84 &16 &24 &24 &28 &28 &50 \\
        & &\textbf{LEAP} &19 &19 &17 &9 &19 &17 &17 &32 &53 &18 &28 &30 &28 &30 &28 \\
        & &\textbf{Qiskit Synthesized} &50 &32 &126 &49 &37 &45 &120 &410 &4169 &785 &851 &851 &851 &850 &851 \\\cmidrule{1-18}
        \multirow{8}{*}{\textbf{Depth}} &\multirow{4}{*}{\textbf{All-to-All}} &\textbf{Qiskit Mapped} &11 &11 &16 &11 &8 &8 &12 &11 &116 &10 &73 &157 &423 &563 &668 \\
        & &\textbf{QFAST} &17 &17 &15 &7 &17 &15 &15 &21 &61 &9 &21 &29 &29 &29 &35 \\
        & &\textbf{LEAP} &15 &13 &13 &7 &13 &11 &13 &19 &39 &13 &21 &24 &19 &27 &21 \\
        & &\textbf{TKet Mapped} &10 &8 &16 &5 &8 &9 &11 &10 &76 &7 &61 &133 &361 &481 &571 \\
        & &\textbf{Qiskit Synthesized} &29 &17 &56 &26 &21 &19 &51 &121 &1062 &227 &421 &421 &421 &421 &421 \\\cmidrule{2-18}
        &\multirow{4}{*}{\textbf{Linear}} &\textbf{Qiskit Mapped} &23 &24 &29 &23 &21 &18 &17 &32 &136 &10 &73 &157 &423 &563 &668 \\
        & &\textbf{QFAST} &17 &17 &15 &9 &17 &15 &17 &63 &63 &9 &13 &21 &25 &21 &31 \\
        & &\textbf{LEAP} &17 &17 &15 &7 &17 &15 &15 &27 &41 &15 &23 &23 &25 &23 &21 \\
        & &\textbf{TKet Mapped} &12 &11 &15 &6 &8 &8 &12 &11 &104 &10 &73 &157 &423 &563 &668 \\
        & &\textbf{Qiskit Synthesized} &56 &34 &139 &55 &38 &51 &132 &390 &3949 &770 &852 &852 &852 &852 &852 \\\cmidrule{1-18}
        \multirow{8}{*}{\textbf{Parallelism}} &\multirow{4}{*}{\textbf{All-to-All}} &\textbf{Qiskit Mapped} &1.64 &1.27 &1.50 &1.36 &1.88 &1.75 &1.42 &1.91 &1.40 &1.30 &1.78 &1.82 &1.84 &1.85 &1.85 \\
        & &\textbf{QFAST} &1.59 &1.59 &1.60 &1.71 &1.59 &1.60 &1.60 &2.33 &2.20 &3.11 &2.19 &1.79 &2.00 &1.59 &1.91 \\
        & &\textbf{LEAP} &1.60 &1.62 &1.62 &1.71 &1.62 &1.64 &1.62 &2.00 &1.82 &1.69 &1.90 &1.71 &2.11 &1.70 &1.90 \\
        & &\textbf{TKet Mapped} &19 &15 &22 &6 &16 &15 &16 &16 &104 &10 &73 &157 &423 &563 &668 \\
        & &\textbf{Qiskit Synthesized} &1.17 &1.18 &1.27 &1.15 &1.33 &1.21 &1.29 &1.27 &1.16 &1.25 &1.14 &1.14 &1.14 &1.14 &1.14 \\\cmidrule{2-18}
        &\multirow{4}{*}{\textbf{Linear}} &\textbf{Qiskit Mapped} &1.52 &1.46 &1.52 &1.43 &1.57 &1.61 &1.53 &1.78 &1.40 &1.30 &1.78 &1.82 &1.84 &1.85 &1.85 \\
        & &\textbf{QFAST} &1.59 &1.59 &1.60 &1.67 &1.59 &1.60 &1.59 &1.78 &1.97 &2.44 &2.62 &1.62 &1.60 &1.90 &2.35 \\
        & &\textbf{LEAP} &1.59 &1.59 &1.60 &1.71 &1.59 &1.60 &1.60 &1.70 &1.88 &1.67 &1.74 &1.87 &1.60 &1.87 &1.90 \\
        & &\textbf{TKet Mapped} &0.01 &0.01 &0.01 &0.01 &0.01 &0.01 &0.01 &0.03 &0.18 &0.01 &0.10 &0.23 &0.66 &0.90 &1.08 \\
        & &\textbf{Qiskit Synthesized} &1.43 &1.44 &1.44 &1.44 &1.47 &1.43 &1.44 &1.68 &1.72 &1.64 &1.61 &1.61 &1.61 &1.61 &1.61 \\\cmidrule{1-18}
        \multirow{8}{*}{\textbf{Time (s)}} &\multirow{4}{*}{\textbf{All-to-All}} &\textbf{Qiskit Mapped} &0.04 &0.04 &0.05 &0.05 &0.04 &0.08 &0.04 &0.05 &0.36 &0.03 &0.20 &0.40 &1.00 &1.33 &1.67 \\
        & &\textbf{QFAST} &1.82 &1.77 &1.82 &0.23 &4.57 &0.54 &0.70 &7.71 &553.79 &1.29 &13.19 &12.26 &10.87 &6.12 &11.29 \\
        & &\textbf{LEAP} &2.99 &1.89 &1.84 &0.47 &1.01 &0.60 &0.98 &34.57 &2006.31 &10.56 &42.59 &16.41 &31.73 &30.71 &51.12 \\
        & &\textbf{TKet Mapped} &0.02 &0.01 &0.01 &0.01 &0.01 &0.01 &0.01 &0.03 &0.19 &0.01 &0.11 &0.25 &0.70 &0.94 &1.14 \\
        & &\textbf{Qiskit Synthesized} &0.26 &0.14 &0.86 &0.24 &0.22 &0.19 &0.60 &1.58 &12.10 &2.85 &3.36 &3.50 &3.37 &3.52 &3.32 \\\cmidrule{2-18}
        &\multirow{4}{*}{\textbf{Linear}} &\textbf{Qiskit Mapped} &0.17 &0.15 &0.17 &0.15 &0.16 &0.18 &0.13 &0.20 &1.04 &0.06 &0.32 &0.66 &1.82 &2.54 &2.93 \\
        & &\textbf{QFAST} &1.66 &1.64 &1.78 &0.41 &1.60 &1.25 &1.89 &16.25 &201.63 &0.64 &1.77 &1.85 &3.00 &2.81 &6.08 \\
        & &\textbf{LEAP} &2.42 &1.62 &1.42 &0.21 &1.52 &1.13 &0.72 &32.61 &765.19 &1.93 &57.15 &18.82 &9.54 &12.80 &11.24 \\
        & &\textbf{TKet Mapped} &0.02 &0.02 &0.03 &0.02 &0.03 &0.02 &0.02 &0.04 &0.27 &0.02 &0.17 &0.38 &1.14 &1.44 &1.72 \\
        & &\textbf{Qiskit Synthesized} &0.45 &0.28 &1.96 &0.46 &0.37 &0.39 &1.13 &4.07 &41.59 &7.55 &7.02 &8.25 &6.44 &8.47 &7.00 \\
        \bottomrule
        \end{tabular}
    }
\end{table*}

\section{Evaluation}


\begin{table}[]
  \caption{\label{tab:summa} \it \footnotesize  Summary of the 
    quality metrics (average value) for five- and six-qubit circuit synthesis. * Qiskit's methods are exact, yet due to some post-processing in their mapping pipeline, large errors are shown.}
    \tiny
  \begin{tabular}{|l|lllll|lllll|}
\cline{2-11}
\multicolumn{1}{l|}{} & \multicolumn{5}{c|}{All-to-all}                                                                                          & \multicolumn{5}{c|}{Linear}                                                                                              \\ \cline{2-11}
\multicolumn{1}{l|}{} & \multicolumn{1}{l|}{Qiskit Mapped} & \multicolumn{1}{l|}{Tket Mapped} & \multicolumn{1}{l|}{LEAP} & \multicolumn{1}{l|}{QFAST} & \multicolumn{1}{l|}{Qiskit Synthesis} & \multicolumn{1}{l|}{Qiskit Mapped} &  \multicolumn{1}{l|}{Tket Mapped} & \multicolumn{1}{l|}{LEAP} & \multicolumn{1}{l|}{QFAST} & \multicolumn{1}{l|}{Qiskit Synthesis} \\ \cline{1-11}
Time (s)              & \textless{}1                & \textless{}1 & 7.34e3                    & 423                        & 31                                & 1.4             &     \textless{}1        & 608                       & 342                        & 76                                \\
Error                 & 1e-16               & 3e-15        & 1e-12                     & 1e-4                       & 1e-11                             & 2.9e-1*           &    3e-15       & 1e-12                     & 1e-5                       & 9e-1*                              \\
CNOT                  & 240             &     240        & 18.85                     & 27.8                       & 1991                              & 250               &    248.6       & 18.8                      & 36.4                       & 6115                              \\
$U_3$                  & 270          & 243.07               & 41.71                     & 60.9                       & 2155                              & 291           &        270.27       & 42.7                      & 78.2                       & 9512                              \\
Depth                 & 207             &     206.67       & 29.2                      & 43.9                       & 3912                              & 321             &      215.47       & 28                        & 48.6                       & 9004\\
\cline{1-11}
\end{tabular}
\end{table}

  Summarized results are presented in Table~\ref{tab:summa}, with more
  details in Tables~\ref{tab:ql4}~and~\ref{tab:567q}.
 We present data for all-to-all and nearest-neighbor chip topology.

\begin{table*}[htbp!]
  \caption{\label{tab:ql4}\it \footnotesize  Summary of synthesis results for
  QSearch and LEAP on the linear topology. LEAP produces very similar results as QSearch in significantly less time.
	}
	\tiny
	\centering
	\begin{tabular}{|c|c|c|ccc|ccc|}
		\hline
		\multicolumn{3}{|c|}{}&\multicolumn{3}{c|}{QSearch}&\multicolumn{3}{c|}{LEAP} \\
		\midrule
		ALG & Qubits &Ref& CNOT  \begin{tikzpicture}[scale=0.3]
		\linett
		\end{tikzpicture}  &
		Unitary Distance &
		Time (s)   & CNOT  \begin{tikzpicture}[scale=0.3]
		\linett
		\end{tikzpicture}  &
		Unitary Distance &
		Time (s)
		\\
		\hline
		QFT & 3 & 6  & 7  & $3.33*10^{-16}$ & 2.0 & 8 & $2.22*10^{-16}$ & 1.7  \\
		Toffoli & 3 & 6 & 8 & $2.22*10^{-16}$ & 3.4 & 8 & $2.22*10^{-16}$ & 1.6  \\
		Fredkin & 3 & 8 & 8 & $4.44*10^{-16}$ & 2.6 & 8 & $3.33*10^{-16}$ & 1.7  \\
		Peres & 3 & 5 & 7 & $ 0 $ & 1.7 & 7 & $2.22*10^{-16}$ & 1.1  \\
		Logical OR & 3 & 6 & 8 & $2.22*10^{-16}$ & 3.4 & 8 & $3.33*10^{-16}$ & 1.6 \\
		\hline
		QFT & 4 & 12 & 14 & $6.7*10^{-16}$ &2429.3 & 13 & $6.7*10^{-16}$ & 77.9 \\
		TFIM-1&	4&	6&	6&	0&	13.4&	6&	0&	7.2\\
		TFIM-10&	4&	60&	11&	$9.08*10^{-11}$&	955.4&	11&	$3.95*10^{-11}$&	47.8\\
		TFIM-22&	4&	126&	12	& $1.22*10^{-15}$	&2450.3	&12&	$7.77^{-16}$&	41.6\\
		TFIM-60&	4&	360&	12	&$ 4.44*10^{-16} $&	1391&	12&$ 2.22*10^{-16} $&	31.6\\
		TFIM-80&	4&	480&	12&$ 4.44*10^{-16} $	&1553.1	&12&$ 2.22*10^{-16} $&	35 \\
		TFIM-95&	4&	570&	12&$ 6.66*10^{-16} $&	1221.4&	12&$ 2.22*10^{-16} $&	38.1\\
		\bottomrule
	\end{tabular}
\end{table*}
  
  Table~\ref{tab:ql4} presents a direct comparison between QSearch and
  LEAP for circuits up to four qubits. Despite its heuristics, LEAP produces optimal depth
  solutions, matching the reference implementations on nearest-neighbor chip
  topology.  Overall, LEAP can compile four-qubit unitaries up to $59\times$ faster than QSearch.

\begin{table*}[hbtp!]\centering
\caption{\it \footnotesize Results for 5--6 qubit synthesis benchmarks with
QFAST, LEAP, and IBM Qiskit. (* implies the program timed out after 12 hours.)
\label{tab:567q}}
\tiny
\resizebox{\columnwidth}{!}{%
    \setlength\tabcolsep{3pt}
    \begin{tabular}{|c|c|c|cccccccccc|ccccc|}\toprule
        \textbf{} &\textbf{} &\textbf{} &\multicolumn{10}{c}{\textbf{5 Qubits}} &\multicolumn{5}{c}{\textbf{6 Qubits}} \\\cmidrule{4-18}
        \textbf{} &\textbf{} &\textbf{} &\textbf{grover5} &\textbf{hlf} &\textbf{mul} &\textbf{qaoa} &\textbf{qft5} &\textbf{TFIM-10} &\textbf{TFIM-40} &\textbf{TFIM-60} &\textbf{TFIM-80} &\textbf{TFIM-100} &\textbf{TFIM-1} &\textbf{TFIM-10} &\textbf{TFIM-24} &\textbf{TFIM-31} &\textbf{TFIM-51} \\\midrule
        \multirow{8}{*}{\textbf{CNOTs}} &\multirow{4}{*}{\textbf{All-to-All}} &\textbf{Qiskit Mapped} &48 &13 &17 &20 &20 &80 &320 &480 &640 &800 &10 &100 &240 &310 &510  \\
        & &\textbf{TKet Mapped} &48 &7 &15 &20 &20 &80 &320 &480 &640 &800 &10 &100 &240 &310 &510 \\
        & &\textbf{LEAP} & * & 9 & 13 & * & 31 & 18 & 22 & 21 & 21 & 22 & 10 & * & * & * & * \\
        & &\textbf{QFAST} &70 &13 &18 &39 &46 &20 &20 &24 &22 &26 &12 &29 &26 &24 &28 \\
        & &\textbf{Qiskit Synthesized} &570 &870 &77 &750 &580 &1025 &1025 &1025 &1025 &1025 &4006 &4474 &4474 &4474 &4474 \\\cmidrule{2-18}
        &\multirow{4}{*}{\textbf{Linear}} &\textbf{Qiskit Mapped} &131 &23 &22 &55 &31 &80 &320 &480 &640 &800 &10 &100 &240 &310 &510  \\
        & &\textbf{TKet Mapped} &96 &16 &24 &42 &41 &80 &320 &480 &640 &800 &10 &100 &240 &310 &510 \\
        & &\textbf{LEAP} & 49 & 15 & 15 & 28 & 30 & 18 & 20 & 20 & 20 & 20 & 10 & 24 & 27 & 29 & 30 \\
        & &\textbf{QFAST} &60 &55 &58 &69 &114 &12 &18 &20 &20 &21 &10 &16 &20 &22 &32 \\
        & &\textbf{Qiskit Synthesized} &2503 &2578 &760 &2692 &2622 &2791 &2791 &2791 &2791 &2791 &13155 &13365 &13365 &13365 &13365 \\\cmidrule{1-18}
        \multirow{6}{*}{$U_3$\textbf{s}} &\multirow{3}{*}{\textbf{All-to-All}} &\textbf{Qiskit Mapped} &78 &8 &16 &20 &29 &90 &360 &540 &720 &900 &11 &110 &264 &341 &561 \\
        & &\textbf{TKet Mapped} &72 &10 &16 &19 &29 &81 &321 &481 &641 &801 &11 &101 &241 &311 &511 \\
        & &\textbf{LEAP} & * & 22 & 27 & * & 65 & 41 & 49 & 45 & 47 & 45 & 26 & * & * & * & * \\
        & &\textbf{QFAST} &145 &31 &41 &83 &97 &45 &45 &53 &49 &57 &30 &64 &58 &54 &62 \\
        & &\textbf{Qiskit Synthesized} &672 &976 &87 &861 &687 &1140 &1140 &1140 &1140 &1140 &4294 &4765 &4765 &4765 &4765 \\\cmidrule{2-18}
        &\multirow{3}{*}{\textbf{Linear}} &\textbf{Qiskit Mapped} &235 &37 &40 &93 &63 &90 &360 &540 &720 &900 &11 &110 &264 &341 &561 \\
        & &\textbf{TKet Mapped} &72 &10 &15 &19 &29 &81 &321 &481 &641 &801 &11 &101 &241 &311 &511 \\
        & &\textbf{LEAP} & 103 & 35 & 35 & 61 & 65 & 41 & 45 & 45 & 45 & 45 & 26 & 54 & 60 & 64 & 64 \\
        & &\textbf{QFAST} &125 &115 &121 &143 &233 &29 &41 &45 &45 &47 &26 &38 &46 &50 &70 \\
        & &\textbf{Qiskit Synthesized} &4008 &4046 &1190 &4264 &4165 &4400 &4401 &4401 &4401 &4400 &20375 &20659 &20658 &20656 &20658 \\\cmidrule{1-18}
        \multirow{6}{*}{\textbf{Depth}} &\multirow{3}{*}{\textbf{All-to-All}} &\textbf{Qiskit Mapped} &85 &16 &26 &32 &26 &76 &286 &426 &566 &706 &16 &79 &177 &226 &366  \\
        & &\textbf{TKet Mapped} &85 &8 &25 &32 &26 &76 &286 &426 &566 &706 &16 &79 &177 &226 &366 \\
        & &\textbf{LEAP} & * & 13 & 22 & * & 47 & 23 & 35 & 31 & 31 & 33 & 13 & * & * & * & * \\
        & &\textbf{QFAST} &123 &21 &33 &65 &85 &31 &33 &49 &29 &39 &13 &47 &29 &29 &33 \\
        & &\textbf{Qiskit Synthesized} &1064 &1662 &138 &1451 &1089 &2008 &2008 &2008 &2008 &2008 &7872 &8841 &8841 &8841 &8841 \\\cmidrule{2-18}
        &\multirow{3}{*}{\textbf{Linear}} &\textbf{Qiskit Mapped} &200 &34 &40 &76 &44 &76 &286 &426 &566 &706 &16 &79 &177 &226 &366 \\
        & &\textbf{TKet Mapped} &133 &17 &36 &53 &43 &76 &286 &426 &566 &706 &16 &79 &177 &226 &366 \\
        & &\textbf{LEAP} & 71 & 17 & 29 & 41 & 45 & 25 & 27 & 35 & 35 & 31 & 13 & 31 & 31 & 33 & 37 \\
        & &\textbf{QFAST} &99 &87 &77 &83 &151 &17 &21 &29 &21 &27 &13 &17 &25 &21 &45 \\
        & &\textbf{Qiskit Synthesized} &3799 &3933 &1115 &4061 &3924 &4236 &4236 &4236 &4236 &4236 &19074 &19495 &19495 &19494 &19494 \\\cmidrule{1-18}
        \multirow{6}{*}{\textbf{Time (s)}} &\multirow{3}{*}{\textbf{All-to-All}} &\textbf{Qiskit Mapped} &0.16 &0.05 &0.07 &0.07 &0.11 &0.22 &0.88 &1.19 &1.68 &2.03 &0.04 &0.28 &0.62 &0.80 &1.41 \\
        & &\textbf{TKet Mapped} &0.07 &0.03 &0.03 &0.04 &0.04 &0.15 &0.68 &1.03 &1.41 &1.79 &0.03 &0.20 &0.46 &0.59 &0.97 \\
        & &\textbf{LEAP} & * & 618.62 & 652.92 & * & 11418.54 & 7826.57 & 16527.44 & 9069.7 & 6628.47 & 1586.35 & 19233.36 & * & * & * & * \\
        & &\textbf{QFAST} &3187.40 &27.70 &86.79 &249.15 &499.49 &79.86 &69.38 &71.98 &77.42 &215.13 &23.14 &618.43 &191.99 &270.70 &684.63 \\
        & &\textbf{Qiskit Synthesized} &11.61 &14.50 &2.65 &14.61 &14.43 &14.35 &15.04 &14.59 &14.27 &16.52 &82.16 &62.93 &64.10 &63.34 &64.62 \\\cmidrule{2-18}
        &\multirow{3}{*}{\textbf{Linear}} &\textbf{Qiskit Mapped} &1.12 &0.24 &0.38 &0.46 &0.34 &0.43 &1.75 &2.57 &3.39 &4.31 &0.09 &0.51 &1.30 &1.61 &2.60 \\
        & &\textbf{TKet Mapped} &0.14 &0.04 &0.05 &0.06 &0.07 &0.25 &1.10 &1.65 &2.12 &2.69 &0.06 &0.34 &0.76 &0.98 &1.58 \\
        & &\textbf{LEAP} & 25233.78 & 165.50 & 856.36 & 3525.54 & 5165.28 & 11631.55 & 3585.95 & 2113.57 & 1901.41 & 2835.3 & 7651.29 & 145303.80 & 175491.42 & 177015.25 & 47681.98 \\
        & &\textbf{QFAST} &992.38 &228.55 &213.94 &365.15 &1901.26 &7.67 &22.78 &26.63 &30.28 &21.01 &5.25 &61.68 &82.52 &408.35 &772.39 \\
        & &\textbf{Qiskit Synthesized} &33.20 &34.42 &12.38 &36.25 &38.37 &35.93 &35.53 &32.27 &34.11 &32.41 &170.08 &161.25 &156.66 &161.30 &159.81 \\
        \bottomrule
        \end{tabular}
}
\end{table*}

As shown in Table~\ref{tab:567q}, LEAP scales up to six qubits. In
this case, we include full topology data, as well results for
compilation with QFAST, Qiskit, Qiskit-synth, and Tket. On well-known quantum
circuits such as VQE and QFT and physical simulation circuits such as 
TFIM, LEAP with re-synthesis can reduce the CNOT count by up to 48×,
or 11× on average when compared to Qiskit. On average when compared to Tket, LEAP
 reduces depth by a factor of 7×. Our heuristics rarely affect solution quality,
 and LEAP can match optimal depth solutions. At five and six qubits, LEAP
synthesizes circuits with to 1.19× fewer CNOTs on average compared with QFAST,
albeit with an average 3.55× performance penalty. LEAP can be one order of
magnitude slower than Qiskit-synth while providing two or more orders of
magnitude shorter circuits.

\subsection{Impact of Prefix Synthesis}

\begin{table*}[htbp!]
	\caption{\label{tab:prefix}\it \footnotesize  Number and location of prefix blocks for various circuits. }
	\tiny
	\centering
	\begin{tabular}{|c|c|c|c|c|}
		\toprule
		ALG & Qubits & CNOT  \begin{tikzpicture}[scale=0.3]
			\linett
		\end{tikzpicture}  & \# of Blocks & Block End Locations \\
	    \midrule
		fredkin&3&8&2&5,8\\
		toffoli&3&8&2&6,9\\
		grover3&3&7&2&5,7\\
		hhl&3&3&1&3,\\
		or&3&8&2&5,8\\
		peres&3&7&2&6,7\\
		qft3&3&8&2&5,9\\
		qft4&4&18&4&5,13,18,21\\
		adder&4&15&3&8,14,19\\
		vqe&5&20&8&3,7,11,14,18,21,25,28\\
		TFIM-1&4&7&2&5,7\\
		TFIM-10&4&12&3&5,10,12\\
		TFIM-22&4&12&3&5,10,12\\
		TFIM-60&4&12&3&5,10,12\\
		TFIM-80&4&12&3&5,10,12\\
		TFIM-95&4&12&3&5,10,12\\
		mul&5&15&5&3,9,12,16,18\\
		qaoa&5&28&7&6,10,14,19,24,29,35\\
		qft5&5&30&10&5,8,11,15,20,25,30,35,38,40\\
		TFIM-10&5&18&7&3,6,9,13,16,19,21\\
		TFIM-40&5&20&7&3,7,10,13,16,19,21\\
		TFIM-60&5&20&7&3,6,10,15,18,21,24\\
		TFIM-80&5&20&7&3,6,11,16,20,23,24\\
		TFIM-100&5&20&6&5,9,13,17,20,22\\
		TFIM-1&6&10&4&4,7,10,12\\
		\hline
	\end{tabular}
\end{table*}

Most of the speed improvements are directly attributable to prefix synthesis,
which reduces by orders of magnitude the number of partial solutions evaluated.
For example, for QFT4, the whole search space contains $\approx 43M$ solution
candidates. QSearch will explore 2,823 nodes, while LEAP will explore 410. For
{\tt TFIM-22}, these numbers are ($\approx 1.6M$, 54,020, 176) respectively.
Detailed
results are omitted for brevity.

Prefix formation is calculated based on a best-fit line formed by a
linear regression of the best scores versus the depth associated with
the new best-found score. This linear regression is used as an estimator of the expected score at the current depth. When the score calculated from the heuristic is better than the expected score, this means that the new best score is better than expected; in other words, more progress to the solution has been made than expected. We note that when the search algorithm in QSearch needs to backtrack and search many different nodes, the progress towards the solution is slower, and the calculated score is worse than the expected score. We, therefore, do not form prefixes in this case, which allows LEAP to maintain the important backtracking and searching that makes QSearch optimal.

 Table~\ref{tab:prefix} presents the
number of prefixes formed during synthesis for each circuit
considered. Since prefixes have a depth between three and five qubits,
this informs our choice of the re-synthesis window discussed below.

\subsection{Impact of Incremental Re-synthesis}
While significantly reducing depth (with respect to the circuit reference),
prefix synthesis can be improved upon by incremental re-synthesis, as shown
by the comparison in Table~\ref{tab:iro}.
LEAP applies only a single step of re-synthesis.  Given the solution from
prefix synthesis, LEAP selects a window at each prefix boundary, resynthesizes,
and reassmembles the circuit. Detailed results are omitted for brevity, but
further iterations do little to improve the solution.

\begin{table*}[htbp!]
	\caption{\label{tab:iro}\it \footnotesize  Summary of the CNOT reduction and time for resynthesis on the linear topology. }
	\tiny
	\centering
	\begin{tabular}{|c|c|ccc|ccc|}
		\toprule
		&&\multicolumn{3}{c|}{Before Resynthesis}&\multicolumn{3}{c|}{After Resynthesis} \\
		\hline
		ALG & Qubits & CNOT  \begin{tikzpicture}[scale=0.3]
		\linett
		\end{tikzpicture}  &
		Unitary Distance &
		Time (s)   & CNOT  \begin{tikzpicture}[scale=0.3]
		\linett
		\end{tikzpicture}  &
		Unitary Distance &
		Time (s)
		\\		
		\hline
		qft3& 3&	9&	0&	1.6&	8&	0&	3.4\\

		logical or& 3&	8&$ 4.44*10^{-16} $&	1.4&	8&$ 4.44*10^{-16} $&	5.9\\

		fredkin& 3&	8&$ 2.22*10^{-16} $&	1.4	&8&$ 2.22*10^{-16} $&	5.7\\
		
		toffoli& 3&	9&$ 2.22*10^{-16} $&	1.7&	8&	0&	3.4\\
		
		adder& 4 &	19&	0&	48.9&	15&$ 2.22*10^{-16} $&	76.7\\
		
		qft4& 4&	21&$ 2.22*10^{-16} $&	38.6&	18&$ 1.11*10^{-16} $&	190.3\\
		
		TFIM-10& 4&	12&$ 8.03*10^{-12} $&	10.3&	12&$ 8.03*10^{-12} $&	176.6\\
		
		TFIM-80& 4&	12&$ 6.66*10^{-16} $&	4.2&	12&$ 6.66*10^{-16} $&	103.8\\
		
		TFIM-95& 4&  12&$ 4.44*10^{-16} $&	6.5&	12&$ 4.44*10^{-16} $&	113\\
		
		vqe& 4& 	28&$ 2.47*10^{-11} $&	151.2&	20&$ 2.70*10^{-11} $&	2062.8\\

		qft5& 5&	40&$ 1.22*10^{-15} $&	772.4&	30&$ 6.66*10^{-16} $&	4392.8\\

		TFIM-10& 5&	21&$ 7.97*10^{-12} $&	310.6&	18&$ 9.19*10^{-12} $&	11320.8\\
		
		TFIM-40& 5&	21&$ 6.66*10^{-16} $&	44&	20&	0&	3541.8\\
		
		TFIM-60& 5&	24&	0&	66.9&	20&	0&	2046.5\\

		TFIM-80& 5&	24&$ 2.22*10^{-16} $&	73.5&	20&$ 2.22*10^{-16} $&	1827.8\\

		TFIM-100& 5 &	22&$ 4.44*10^{-16} $&	55.4&	20&$ 1.11*10^{-16} $&	2779.8\\

		mul& 5&	18&$ 4.44*10^{-16} $&	47.0&	15&$ 2.22*10^{-16} $&	809.2\\
		
		TFIM-1& 6&	12&$ 2.22*10^{-16} $&	213.3&	10&$ 1.11*10^{-16} $&	7437.9\\
		
		\bottomrule
	\end{tabular}
\end{table*}

The re-synthesis window in LEAP is chosen pragmatically with a limited depth
(7 CNOTs for 3 and 4 qubits, 5 CNOTs for 5 and 6 qubits in our case), to lead to reasonable expectations on execution
time, while providing some optimization potential.

Incremental re-synthesis reduces circuit depth by 15\% on average,
albeit  in many cases with a significant impact on the runtime.

\subsection{Impact of Dimensionality Reduction}

LEAP applies a single  step of dimensionality reduction at the end of the
synthesis process, the sweep  starting at the circuit 
beginning. For brevity, we omit detailed data and note that in
this final stage dimensionality reduction eliminates up to 40\% of $U_3$
gates (parameters) and shortens the circuit critical path. These results
indicate that our approach overfits the problem by inserting
too many $U_3$ gates.

\begin{table}[htbp!]
  \caption{\label{fig:delu3} \it \footnotesize  Spatial placement of
    $U_3$ gates deleted. The number of columns denotes circuit stages (CNOTs),
  and we present the number of gates deleted at each position.}
 \tiny
 \centering
 \begin{tabular}{|c|c|c|c|c|c|c|c|c|c|}
 	\toprule
 	Name&\multicolumn{9}{c|}{Number of Gates Deleted} \\
 	\hline
 	qft2&2&0&0&0&&&&&\\
 	qft3&2&0&0&1&0&0&1&1& \\
 	fredkin&3&2&0&1&1&2&0&0&1\\
 	toffoli&2&2&1&2&1&2&0&1&0\\
 	peres&2&0&1&2&0&1&0&1&\\
 	logical\_or&2&1&2&0&2&1&0&1&0\\
 	hhl&2&0&2&0&&&&&\\
 	\bottomrule
 \end{tabular}
\end{table}

We examined the spatial occurrence of single-qubit
gate deletion since this may guide any dynamic attempts to eliminate
parameters during  synthesis for scalability
purposes. Figure~\ref{fig:delu3} presents a summary for three-qubit
circuits; trends are similar for all other benchmarks considered. The
data shows that gate deletion is successful at many  circuit layers, indicating
that a heuristic for on-the-fly dimensionality reduction heuristic may
be feasible to develop  for even  further scalability and quality improvements. As
discussed in Section~\ref{sec:ms}, dimensionality reduction will
reduce the number of parameters for numerical optimization, while
reducing overfitting and gate (parameter) correlation that lead to
cancellations of gate effects on a qubit.

\subsection{Impact of Multistart Optimization}

\begin{table*}
	\caption{\label{tab:multi}\it \footnotesize  Accuracy and speed of various optimizers on a variety of circuits. APOSMM-N means APOSMM with $N$ starting points. }
	\tiny
	\centering
	\resizebox{\columnwidth}{!}{%
	\begin{tabular}{|c|c|cc|cc|cc|cc|cc|cc|cc|}
		\toprule
		&&\multicolumn{2}{c|}{BFGS}&\multicolumn{2}{c|}{Ceres}&\multicolumn{2}{c|}{APOSMM-8}&\multicolumn{2}{c|}{APOSMM-12}&\multicolumn{2}{c|}{APOSMM-16}&\multicolumn{2}{c|}{APOSMM-20}&\multicolumn{2}{c|}{APOSMM-24} \\
		\hline
		ALG & CNOT  \begin{tikzpicture}[scale=0.3]
			\linett
		\end{tikzpicture}  &$\%$ Success & Time (s) &$\%$ Success & Time (s) &$\%$ Success & Time (s) &$\%$ Success & Time (s) &$\%$ Success & Time (s) &$\%$ Success & Time (s) &$\%$ Success & Time (s)
		\\		
		\hline
		fredkin&8&89&0.03&69&0.01&100&0.13&100&0.14&100&0.14&100&0.15&100&0.16 \\ logical\_or&8&16&\scalebox{.75}{$<$} 0.01&55&0.01&100&0.13&100&0.14&100&0.15&100&0.16&100&0.17 \\ peres&7&18&\scalebox{.75}{$<$} 0.01&73&0.01&69&0.08&90&0.11&92&0.12&98&0.13&99&0.14 \\
		toffoli&8&43&0.01&74&0.01&100&0.13&100&0.14&100&0.14&100&0.15&100&0.17 \\ qft3&8&9&\scalebox{.75}{$<$} 0.01&26&\scalebox{.75}{$<$} 0.01&80&0.10&91&0.12&95&0.13&98&0.14&100&0.16 \\
		qft4&18&1&\scalebox{.75}{$<$} 0.01&15&0.02&66&0.50&83&0.68&92&0.82&94&0.99&99&1.08 \\
		qft5&30&0&\scalebox{.75}{$<$} 0.01&2&0.12&8&1.19&13&2.78&15&3.81&25&7.21&36&12.10 \\ 
		\bottomrule
	\end{tabular}
	}
\end{table*}

When evaluating numerical optimizers used in
synthesis, we are interested  in determining how often they found the
true minimum, since this has a significant impact on both solution
quality and execution speed. We evaluated
the commonly used local optimization methods Google's Ceres~\citep{ceres} and an
implementation of L-BFGS~\citep{lbfgs} as well as the multistart
APOSMM~\citep{APOSMM} framework.

We ran each optimizer 100 times on several circuits to evaluate their accuracy
and speed. The results are summarized in Table~\ref{tab:multi}. The QFT results
illustrate that the BFGS and Ceres optimizers perform poorly even on a smaller
circuit such as a three-qubit QFT, finding solutions just 9\% and 26\% of the
time, much lower than even APOSMM with 8 starting points. We found that APOSMM
with 12 starting points performed well on all but the five-qubit QFT circuit.
Since optimizing the parameters of the QFT5 circuit is a much
higher-dimensional problem, even APOSMM with 24 starting points found solutions
in only 36\% of the runs.

While APOSMM is much more accurate than BFGS and Ceres on the circuits we
tested, it is also about an order of magnitude slower for larger circuits, even
though the local optimization runs are done in parallel. In addition, the
slowdown increases with the number of starting points. The time for QFT5
approximately doubles every 4 additional starting points for parallel runs. For
our runs in Table ~\ref{tab:567q} we selected 12 starting points since this
number was reasonably accurate and takes a reasonable amount of time.

Therefore when using LEAP, we use Ceres
because it is fast and scales well, and a missed solution will be found during
re-synthesis. During re-synthesis, APOSMM is used, since it is much more
likely to find true minima, thus strengthening the optimality of search-based
algorithms.

\subsection{Gate Set Exploration}

Similar to QSearch, LEAP can target different native gate sets and
provide another dimension to circuit optimization or hardware design exploration. Besides CNOT, we
have targeted  other two-qubit gates supported by QPU
manufacturers: CSX ($\sqrt{CNOT}$), iSWAP, and SQISW ($\sqrt{iSWAP}$). Here,
the square root gates implement the matrix square root of their
counterpart, and their composition has been previously studied~\citep{sqrt} for
generic two-qubit programs.  Results are presented in
Table~\ref{tab:iswap}. We make the following observations:
\begin{itemize}
\item  While CNOT and iSWAP are considered ``equivalent'' in terms of
expressive power,  using CNOT gates  for larger circuits (five and six qubits)
tends  to produce observably shorter circuits.
\item Mixing two-qubit gates (CNOT+iSWAP) tends to
  produce shorter circuits than when using CNOT alone.
\item The depths of CNOT- and  $\sqrt{CNOT}$-based circuits are very
  similar. Given that in some implementations the latency of
  $\sqrt{CNOT}$ gates may
  be shorter than that of CNOT gates, the former may be able to
  provide a performance advantage.
\item Sleator and Weinfurter~\citep{swt}  prove that the Toffoli gate can be
  optimally  implemented  using a five-gate combination of CNOT and
  $\sqrt{CNOT}$. LEAP can reproduce this result, which
  indicates it may provide a useful tool for discovering optimal
  implementations of previously proposed gates.  
\end{itemize}

These observations are somewhat surprising and probably worth a more
detailed future investigation. While the data indicates that mixing
CNOT and iSWAP can produce the shortest circuits, we found that in
LEAP the search space size would double, hence the speed to the solution
will suffer. Therefore for our experiments, we kept with the CNOT+$U_3$ gate set that was used by QFAST and Qsearch.

\begin{table*}[t]
	\caption{\label{tab:iswap}\it \footnotesize  Number of two
			qubit gates needed to implement various three- to six-qubit circuits. Using CNOT
			 reduces the number of two-qubit gates needed vs iSWAP, whereas a combination of
			  CNOT and iSWAP reduces the number of two-qubit gates even further.
		}
	\tiny
	\centering
	\begin{tabular}{|c|c|c|c|c|c|c|c|}
	\toprule
	ALG & CNOT & SQCNOT & iSWAP & SQISW & CNOT + iSWAP & CNOT + SQCNOT & iSWAP + SQISW \\
	\midrule
	qft3 & 6 & 8 & 7 & 8 & 5 & 5 & 7\\
	fredkin & 7 & 9 & 7 & 9 & 7 & 7 & 8\\
	toffoli & 6 & 7 & 7 & 8 & 6 & 5 & 7\\
	peres & 5 & 5 & 7 & 8 & 5 & 4 & 6\\
	logical or & 6 & 7 & 7 & 8 & 6 & 8 & 7\\
	\hline
	\end{tabular}
	\quad
	\begin{tabular}{|c|c|c|}
	\toprule
	ALG & iSWAP & CNOT \\
	\midrule
	qft4 & 22 & 13 \\
	tfim-4-22 & 16 & 12 \\
	tfim-4-95 & 14 & 12 \\
	vqe & 26 & 21 \\
	full adder & 30 & 18 \\
	hlf & 22 & 13 \\
	mul & 18 & 13 \\
	qft5 & 50 & 28 \\
	tfim-5-40 & 29 & 20 \\
	tfim-5-100 & 33 & 20 \\
	tfim-6-24 & 40 & 28 \\
	tfim-6-51 & 43 & 31 \\
	\hline
	\end{tabular}
\end{table*}

\section{Discussion}
\label{sec:disc}

Overall, the results indicate that
the heuristics employed in LEAP 
are much faster than QSearch and
are still able to produce low-depth
solutions in a topology-aware manner. The average depth difference for  three- and four-qubit
benchmarks between QSearch and LEAP is $0$ across physical chip
topologies and workload.

We find the prefix formation idea intuitive, easily generalizable, and
powerful. The method used to derive prefix formation employs concepts
encountered in numerical optimization algorithms and
is easily identifiable in other search-based synthesis algorithms:
``progress'' to the solution, and ``region of similarity'' or plateau. 

The LEAP algorithm indicates that incremental and iterative approaches
to synthesis work well.  In our case, the results even indicate that
one extra step of local optimization can match the efficacy of
global optimization. This result bodes well for approaches that scale
synthesis past hundreds of qubits through circuit partitioning, such as our
QGo~\citep{wu2020optimizing} optimization and
QuEst~\citep{patel2021robust} approximation algorithms.

Dimensionality reduction as implemented in LEAP not only reduces the effects of overfitting
by numerical optimization but also opens a promising path for scaling
numerical-optimization-based synthesis. Since we were able to delete 40\%
of parameters from the final solution, we believe that by combining
it with prefix synthesis we can further improve LEAP's scalability.

Multistart optimization can be trivially incorporated into any algorithm, and we
have indeed already modified the QSearch and QFAST algorithms to incorporate
it.  Furthermore, the spirit of the multistart ``approach'' can be
employed to further prune the synthesis search space. Whenever a
prefix formed, the synthesis algorithm had explored a plateau and a
local minimum. At this stage, a multistart search could be started using
as seeds other promising partial solutions within the tree.

\begin{figure}[htbp!]
 \comment{ [width=3in,height=1in]}
 \centerline{\includegraphics[width=5.5in,keepaspectratio]{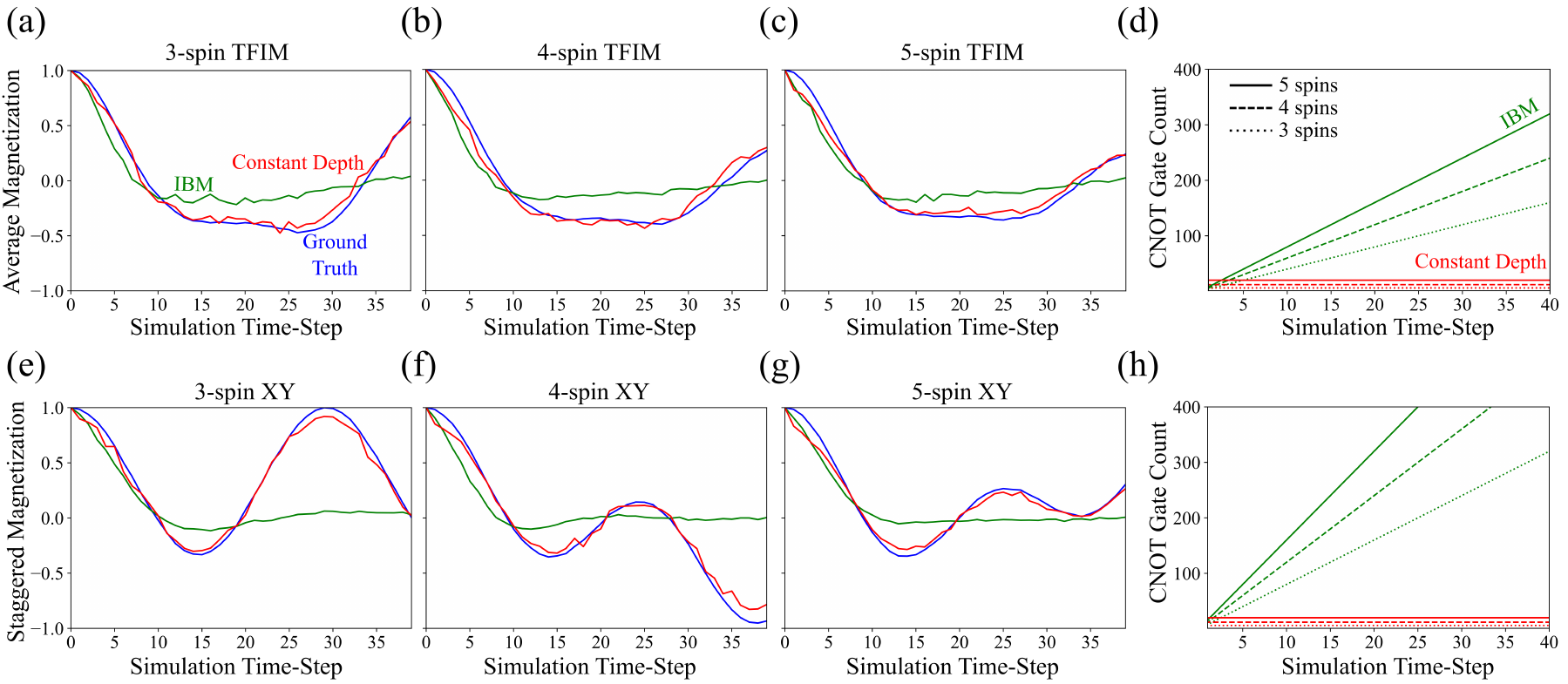}}
  \caption{\label{fig:roel} \it \footnotesize  TFIM circuit depth
    evolution and ``fidelity'' when executed on the IBM Athens system.
    ``IBM'' is compiled with Qiskit, while ``Constant Depth'' is synthesized with LEAP}
\end{figure}

The prefix formation idea is powerful and showcases how synthesis can
turn into a capability tool. TFIM circuits simulate a  time-dependent
Hamiltonian, where the circuit for each time step ``contains'' the
circuit (computation)
associated with the previous time step as a prefix. The circuits
generated by the TFIM domain generator grow linearly in size.  In our experiments,
we observed that after some initial time steps, all circuits  for any
late time step have an
asymptotic constant depth. This observation led to the following experiment: we picked a
circuit structure generated for a late simulation step and considered it as a
parameterized template for all other simulation steps. We then
successfully solved
the numerical optimization problem with this template for any TFIM step.
This procedure
empirically provides us with a fixed-depth (short-depth) template for the TFIM
algorithm. Furthermore, this demonstration motivated a
successful effort~\citep{roel}  to derive from first principles a
fixed-depth circuit for TFIM.
The results are presented in Figure~\ref{fig:roel}. Note the highly
increased fidelity when running the circuit on the IBM Athens system.

The QITE algorithm presents an interesting challenge to the prefix
formation idea. In this case, the next timestep circuit is obtained
by extending  the ``current''  circuit with a block dependent on its
output after execution. When executing on hardware, synthesis has 
real-time constraints, and it has to deal with the hardware noise that
affects the output. Preliminary results, courtesy of our collaborators
Jean-Loup Ville and Alexis Morvan, indicate that the approach taken for
TFIM may be successful for QITE. Table~\ref{tab:qite} summarizes the
preliminary observations and indicates  that again synthesis produces
better-quality circuits than the  domain generator or traditional
compilation does. Note that in this experiment LEAP was fast enough to
produce real-time results during  the hardware experiment only for three-qubit
circuits.

\begin{table}[!htp]\centering
\caption{\label{tab:qite} \it \footnotesize Summary of  QITE results
  when running synthesis on hardware experiments. Structure of any
  circuit is determined by the output of the previous circuit, hence
  hardware noise.}
\tiny
\begin{tabular}{|c|c|c|c|}
  \hline
  & \multicolumn{3}{c|}{\cn} \\
  \hline
     QITE size&   Qiskit Isometry & QFAST & LEAP \\
         
\hline
  2 & 3 & 3&3 \\
     3 & 30-35& 10-12 & 7-12 \\
     4 & 160-200 & 70-80 & 30-50 \\
 \hline
     \end{tabular}

\end{table}

Looking forward, the question remains whether numerical-optimization-based
synthesis can be useful in fault-tolerant quantum computing. There, the
single-qubit gates  will change to Cliffords and the T gate, or another non-Clifford
gate that makes the gate set universal. The execution cost model is also
expected to be different: CNOTs and Cliffords become cheap, while the non-Clifford
operations become expensive.  Likely, the 
non-Cliffords are qualitatively more
``expensive'' than CNOTs in NISQ computing. Thus, the optimization objective becomes
minimizing the number of non-Clifford gates.

We have already shown that LEAP can be  retargeted to new gate
sets. We also have very strong evidence
that adding a multi-objective optimization approach to search-based
synthesis works very well under a fault-tolerant quantum computing
cost model. The data indicates that it is realistic to expect efficacy
improvements similar to those provided by LEAP under the NISQ cost model. This work is ongoing (and due
to intellectual property concerns, we cannot disclose more
details). As the already mentioned scalable partitioning approaches only leverage LEAP and do not require
additional cost models,  this bodes very well for the future of
numerical-optimization-based synthesis in fault-tolerant quantum computing.

\section{Related Work}
\label{sec:related}

A fundamental result that
spurred the apparition of quantum circuit synthesis is provided by the
Solovay--Kitaev (SK) theorem.  The theorem
relates circuit depth to the quality of the approximation, and its
proof is by construction~\citep{DawsonNielson05,Nagy16,ola15}. Different approaches~\citep{DawsonNielson05,ZXZ16,BocharovPRL12,MIM13,Qcompile16,ctmq,23gates,householderQ,CSD04,amy16,seroussi80} to synthesis have been
introduced since, with the goal of generating shorter-depth circuits. 
These can be coarsely classified based on several
criteria:  target gate set, algorithmic approach, and solution distinguishability.

\parah{Target Gate Set} The SK algorithm is applicable to any 
universal gate set. Later examples include
synthesis of z-rotation unitaries with
Clifford+V approximation~\citep{Ross15} or Clifford+T gates~\citep{KMM16}. When ancillary qubits are allowed, one can synthesize
single-qubit unitaries  with the Clifford+T gate set~\citep{KMM16,KSV02,Paetznick2014}. 
While these efforts propelled the field of synthesis, they are not 
used on NISQ devices, which offer a different gate set
($R_x, R_z,CNOT,iSWAP$ and M\o lmer--S\o rensen all-to-all).
Several~\citep{raban,synthcsd,ionsynth}  other
algorithms, discussed below, have since emerged.

\comment{ For example, the z-rotation unitaries can be
synthesized with Clifford+V approximation~\citep{Ross15} or  with Clifford+T gates~\citep{KMM16}. The set of Pauli, Hadamard, Phase, CNOT 
gates form what is known as the Clifford group gates. When augmented with the T gate defined
as 
\[
T = \left(
\begin{array}{cc}
1 & 0 \\
0 & \zeta_8 
\end{array}
\right), \ \ \mbox{where} \ \ \zeta_8 = e^{i\pi/4},
\]
the gate set is universal.
t lead to better complexity $\mathcal{O}(\log^{1.75}(1/\epsilon))$ compared 
to the SK Algorithm.  \comment{$\mathcal{O}(\log(1/\epsilon))$ T-count scaling.}
}  

\parah{Algorithmic Approaches} The early attempts inspired by
the Solovay--Kitaev algorithm use a recursive (or divide-and-conquer) 
formulation, sometimes supplemented with search heuristics at the
bottom. More recent search-based approaches are illustrated by the
meet-in-the-middle~\citep{MIM13}  algorithm.

Several  approaches use techniques from linear algebra for
unitary and tensor decomposition. Bullock and Markov~\citep{23gates} use QR matrix factorization via a Givens rotation and Householder transformation~\citep{householderQ},
but open questions remain as to the suitability
for hardware implementation because these algorithms are expressed 
in terms of row and column updates of a matrix rather than in terms of qubits.

The state-of-the-art upper bounds on circuit depth are provided by
techniques~\citep{synthcsd,raban} that use cosine-sine
decomposition. The cosine-sine decomposition was first
used in~\citep{tucci}  for compilation purposes. In practice,
commercial compilers ubiquitously deploy only 
KAK~\citep{tucci2005kak} decompositions for 2-qubit unitaries.

The basic formulation of these techniques is topology
independent. Specializing for topology increases the upper bound on circuit depth by large constants; Shende et al.~\citep{synthcsd} mention a factor of 9,
improved by Iten et al.~\citep{raban} to $4\times$. 
The published approaches are hard to extend to different qubit gate
sets, however, and it remains to be seen whether they can handle qutrits.\footnote{~\citep{qtsynth} describes a method
using Givens rotations and Householder decomposition.} 

\comment{
There are not many studies published about synthesis of qutrit based
circuits and qutrit gate sets.~\citep{qtsynth} describes a method
using Givens rotations and Householder decomposition. As techniques
for qubit based systems using a similar approach have been
proposed~\citep{23gates}, they may allow an easier combination of
qutrit and qubit based synthesis. }

Several techniques use numerical optimization, much as we did. They
describe the gates in their variational/continuous representation and
use optimizers and search to find a gate decomposition and
instantiation.
The work closest to ours is that of Martinez et al.~\citep{ionsynth}, who  use
numerical optimization and brute-force search to synthesize circuits
for a processor using trapped-ion qubits. Their main advantage is the
existence of all-to-all M\o lmer--S\o rensen gates, which allow a topology-independent approach.
The main difference between our work and theirs
is that they use randomization and genetic algorithms to search the
solution space, while we show a more regimented way.
When Martinez et al. describe their results,
they claim that M\o lmer--S\o rensen counts are directly comparable to CNOT
counts. By this metric, we seem to generate circuits comparable to or shorter
than theirs.  It is not clear how their approach behaves when
topology constraints are present. The direct comparison is further limited 
by the fact that they consider only randomly generated unitaries, rather
than algorithms or well-understood gates such as Toffoli or Fredkin.

Another topology-independent numerical optimization technique is
presented in~\citep{qaqc}. The main contribution is to
use a quantum annealer to do searches over sequences of increasing
gate depth. The authors report results only for two-qubit circuits.

All existing studies focus on the quality of the solution, rather than
synthesis speed. They also report results for low-qubit concurrency:
Khatri et al.~\citep{qaqc} for two-qubit systems, Martinez et al.~\citep{ionsynth} for systems up to
four qubits.

\parah{Solution Distinguishability} 
Synthesis algorithms can be classified as exact or approximate based on
distinguishability.  This is a subtle classification criterion, since many
algorithms can be viewed as either.  For example,
the divide-and-conquer algorithm Meet-in-the-Middle
proposed in~\citep{MIM13},
although designed for exact circuit synthesis, 
may also be used to construct an $\epsilon$-approximate circuit. The results seem to indicate that the algorithm failed
to synthesize a three-qubit QFT circuit.  We classify our implementation as approximate since we rely on numerical optimization and therefore must accept solutions at a small distance from the original
unitary.

\comment{
It allows one to search for
circuits of depth $l$ by only generating circuits of depth at most $\lceil l/2 \rceil$ at the complexity of $\mathcal{O}(|\mathcal{V}_{n,\mathcal{G}}|^{\lceil l/2\rceil}\log |\mathcal{V}_{n,\mathcal{G}}|^{\lceil l/2 \rceil})$, 
where $\mathcal{V}_{n,\mathcal{G}}$ denotes the set of unitaries for depth-one 
$n$-qubit circuit. The MIM algorithm is flexible and allows weights to be
added to the gate set to account for the possibility that some gates,
such as those that do not belong to the Clifford group, may be more expensive
to implement. It also allows ancillas to be used in the synthesis.  The 
algorithm uses a number of heuristics to prune the search tree.  Although
it was originally designed for exact circuit synthesis, the algorithm
may also be used to construct an $\epsilon$-approximate circuit.
}

\section{Conclusion}\label{sec:conc}

In this paper we describe the LEAP compiler and  modifications to a
search and numerical-optimization-based synthesis algorithm.  The
results indicate that we can empirically provide optimal-depth
circuits in a topology-aware manner for programs up to six qubits. The
techniques employed prefix formation,  incremental
re-synthesis, dimensionality reduction, and multistart
optimization and can be easily generalized to other algorithms from this
class.
We believe LEAP provides the best-quality optimizer currently
available for circuits up to six qubits on NISQ hardware.
Furthermore, LEAP is the linchpin in our scalable
synthesis algorithms (QGo~\citep{wu2020optimizing}, QuEst~\citep{patel2021robust}) using
circuit partitioning techniques. With these algorithms, we have
demonstrated the synthesis of circuits up to hundreds of qubits. 
LEAP has been released as part of the BQSkit (Berkeley Quantum
Synthesis Toolkit) infrastructure.

\section*{Acknowledgments} This work was supported by the Advanced Quantum Testbed program and the
Quantum Algorithm Teams program of the Advanced Scientific Computing Research
for Basic Energy Sciences program, Office of Science of the U.S.~Department of
Energy under Contract No.~DE-AC02-05CH11231 and DE-AC02-06CH11357.

\bibliographystyle{unsrtnat}
\bibliography{quantum,bibliography,quant_chem}

\framebox{\parbox{.92\linewidth}{The submitted manuscript has been created by
UChicago Argonne, LLC, Operator of Argonne National Laboratory (``Argonne'').
Argonne, a U.S.\ Department of Energy Office of Science laboratory, is operated
under Contract No.\ DE-AC02-06CH11357.  The U.S.\ Government retains for itself,
and others acting on its behalf, a paid-up nonexclusive, irrevocable worldwide
license in said article to reproduce, prepare derivative works, distribute
copies to the public, and perform publicly and display publicly, by or on
behalf of the Government.  The Department of Energy will provide public access
to these results of federally sponsored research in accordance with the DOE
Public Access Plan \url{http://energy.gov/downloads/doe-public-access-plan}.}}

\end{document}